\documentclass[a4paper,aps,pre,superscriptaddress,floatfix,nofootinbib,twocolumn,longbibliography]{revtex4-1}

\usepackage{bbm}
\usepackage{bbold}
\usepackage{bbm}
\usepackage[pdftex]{graphicx}
\usepackage{latexsym,amsmath,verbatim,amssymb,txfonts}
\usepackage{color}
\usepackage{rotating}
\usepackage{verbatim}
\usepackage{multirow}
\usepackage[english]{babel}
\usepackage{comment}
\usepackage{bm}
\usepackage[hidelinks]{hyperref}

\begin{document}
\def\be{\begin{equation}}
\def\ee{\end{equation}}

\def\bc{\begin{center}}
\def\ec{\end{center}}
\def\bea{\begin{eqnarray}}
\def\eea{\end{eqnarray}}
\newcommand{\updownarrows}{\mathbin\uparrow\hspace{+.0em}\downarrow}
\newcommand{\downuparrows}{\mathbin\downarrow\hspace{-.5em}\uparrow}
\newcommand{\avg}[1]{\langle{#1}\rangle}
\newcommand{\Avg}[1]{\left\langle{#1}\right\rangle}
\newcommand{\mi}{\mathrm{i}} 
\newcommand{\di}{i}    
\def\ie{\textit{i.e.}}
\def\etal{\textit{et al.}}
\def\m{\vec{m}}
\def\G{\mathcal{G}}

\newcommand{\gin}[1]{{\bf\color{blue}#1}}

\title{Weighted simplicial complexes and their representation power\\ of higher-order network data and topology}

\author{Federica Baccini}

\affiliation{Department of Computer Science, University of Pisa, and IIT-CNR of Pisa, Italy}
\affiliation{Institute for Informatics and Telematics, CNR, Pisa, Italy}

\author{Filippo Geraci}
\affiliation{Institute for Informatics and Telematics, CNR, Pisa, Italy}

\author{Ginestra Bianconi}
\affiliation{School of Mathematical Sciences, Queen Mary University of London, London, E1 4NS, United Kingdom}
\affiliation{The Alan Turing Institute, The British Library, London NW1 2DB, United Kingdom}

\begin{abstract}
Hypergraphs and simplical complexes both capture the higher-order interactions of complex systems, ranging from higher-order collaboration networks to brain networks. One open problem in the field is what should drive the choice of the adopted mathematical framework to describe higher-order networks starting from data of higher-order interactions. Unweighted simplicial complexes typically involve a loss of information of the data, though having the benefit to capture the higher-order topology of the data. In this work we show that weighted simplicial complexes allow to circumvent all the limitations of unweighted simplicial complexes to represent higher-order interactions. In particular, weighted simplicial complexes can represent higher-order networks without loss of information, allowing at the same time to capture the weighted topology of the data. The higher-order topology is probed by studying the spectral properties of suitably defined weighted Hodge Laplacians displaying a normalized spectrum. The higher-order spectrum of (weighted) normalized Hodge Laplacians is here  studied  combining cohomology theory with information theory.  In the proposed framework, we quantify and compare the information content of higher-order spectra of different dimension using higher-order spectral entropies and spectral relative entropies. The proposed methodology is  tested on  real higher-order collaboration networks and on the weighted version of the simplicial complex model ``Network Geometry with Flavor".
\end{abstract}

\maketitle

\section{Introduction}

Higher-order networks \cite{bianconi2021higher,battiston2021physics,battiston2020networks,majhi2022dynamics,torres2021and,giusti2016two,salnikov2018simplicial,otter2017roadmap,battiston2022higher,bick2021higher} capture the higher-order interactions of complex systems, including collaboration networks, face-to-face social interaction networks, brain networks, and chemical reaction networks.
For instance, in a collaboration network among scientists, higher-order networks allow to capture interactions of a team of co-authors formed by two or more scientists \cite{patania2017shape,carstens2013persistent}.
Higher-order networks include hypegraphs and simplicial complexes.
Hypergraphs are formed by a set of hyperedges, describing higher-order interactions. Simplicial complexes are formed by simplices, where an $n$-simplex is formed by $n+1$ nodes.
In the literature there is an increasing attention in determining the differences between these two mathematical frameworks, both tailored to model higher-order data.
The difference between unweighted simplicial complexes and hypergraphs is that hypergraphs are arbitrary set of hyperedges, while simplicial complexes are sets of simplices closed under the inclusion of the faces of each simplex in the simplicial complex. This additional property implies, for instance, that if a three-way interaction ($2$-simplex, or filled triangle) $[A,B,C]$ among the nodes $A,B$ and $C$ is present in the simplicial complex, then we must include in the simplicial complex also all the links (pairwise interactions)  and all the nodes which constitute the faces of the triangle, i.e. we should also include the simplices $[A,B],[B,C],[A,C],[A],[B],[C]$.
This can be perceived as a limitation in some application domains, such as in collaboration networks. Indeed, if three authors in a collaboration network have co-authored a paper together, it is not generally the case that also two-author papers written by each pair of scientist in the triangle exist.
On the other hand, simplicial complexes provides the network scientists with very powerful tools coming from algebraic topology \cite{lim2020hodge,josthorak2013spectra,hatcher2005algebraic,bianconi2021higher} to characterize the structure of higher-order datasets \cite{krishnagopal2021spectral,kartun2019beyond,giusti2016two,otter2017roadmap,petri2014homological,petri2013topological,reimann2017cliques,jiang2011statistical,sreejith2016forman}, and the interplay between topology and dynamics \cite{bianconi2021higher,millan2020explosive,barbarossa2020topological,millan2021local,torres2020simplicial,calmon2021topological,bianconi2021topological,arnaudon2021connecting,deville2021consensus,ziegler2022balanced,reitz2020higher,giambagli2022diffusion,schaub2020random,schaub2021signal,sardellitti2021topological,ghorbanchian2021higher,zhao2011entropy,sun2022triadic,taylor2015topological,roddenberry2019hodgenet,hajij2020cell,ebli2020simplicial,bodnar2021weisfeiler,katifori2010damage,rocks2021hidden,faskowitz2022edges,witthaut2022collective,tang2022optimizing,chutani2021hysteresis}.
One direction to solve this dichotomy between hypergraphs and simplicial complexes is to introduce the notion of algebraic topology to treat hypergraphs \cite{jost2019hypergraph,mulas2020coupled}.
Here we pursue another direction and we propose to study {\em weighted} simplicial complexes, which are attracting increasing attention \cite{kelin018weighted,kelinbio2020weighted,ren2018weighted}, where each simplex of the simplicial complex is associated to a real number called {\em weight}. Weighted simplicial complexes retain the property that they are closed under the inclusion of the faces of each simplex. However, in this paper we show that if the weights of the simplices are defined according to our algorithm, it is possible to distinguish between simplices that are only included in the simplicial complex for the closure condition to be satisfied (and do not describe {\em bare} higher-order interactions present in the data), and simplices that are also encoding for {\em bare higher-order interactions}. Therefore, with the proposed choice of weights, simplicial complexes can be used interchangeably to hypergraphs, as they can retain all the information present in the data. 
Moreover, we show that the proposed choice of weights for the weighted simplicial complexes also allows to use algebraic topology of weighted simplicial complexes, and hence to investigate their higher-order topology. 
Indeed, the proposed choice of weights of the simplices allows us to define normalized Hodge Laplacians of every dimension. Normalized Hodge Laplacians are particularly useful to compare the spectral properties of a simplicial complex at different dimension, revealing important aspects of its higher-order structure.
Here we show how the higher-order spectral entropies, that generalize the notion of spectral or Von Neumann entropy of networks \cite{anand2009entropy,anand2011shannon,de2015structural,de2016spectral,ghavasieh2020statistical,ghavasieh2022statistical,villegas2022path}, can be used for characterizing the properties of higher-order diffusion processes \cite{torres2020simplicial,reitz2020higher,ziegler2022balanced,millan2021local,mulder2018network} and their associated  characteristic time-scales.
These theoretical insights have been applied to a real dataset of higher-order  scientific collaboration network,and to  the weighted simplicial complex model ``Network Geometry with Flavor" \cite{bianconi2017emergent,bianconi2016network,mulder2018network,courtney2017weighted,bianconi2015complex} revealing the information content encoded in these higher-order network structures. 
Importantly, when analysing the higher-order collaboration network, we also propose a way to quantify the bare weights associated to each team of collaborators, thus extending to higher-order networks the popular choice of network weights for scientific collaboration networks proposed by Newman in  Ref. \cite{newman2001scientific}. 

Note that in this paper our focus is  establishing how weighted simplicial complexes can be used to capture real data without loss of information. Therefore our approach is different in nature and scope with respect to other recent works \cite{zhang2022higher} aimed at exploring the dynamical effects emerging when  different higher-order representations are considered.

The paper is organized as follows. In Sec. II we discuss weighted simplicial complexes and our proposed choice of the topological weights. In Sec. III we introduce the fundamental aspects of algebraic topology that lead to the definition of higher-order normalized Hodge Laplacians and of normalized Dirac operator. In Sec. IV we discuss the higher-order spectral entropies of simplicial complexes and their properties. In Sec. V and VI we present the application of the proposed mathematical framework to a real collaboration network and to the model ``Network Geometry with Flavor", respectively. Finally, in Sec. VII we provide some concluding remarks. The paper is enriched by an Appendix providing the proof that the proposed Hodge Laplacians are normalized at every order.

\section{Weighted simplicial complexes}

A simplicial complex $\mathcal{K}$ is a type of higher-order network \cite{bianconi2021higher} that is increasingly used to study the underlying topology of data. A simplicial complex encodes the higher-order interactions of complex systems, i.e. the interactions between two ore more nodes. In other words, simplicial complexes allow to go beyond the network description of complex systems based exclusively on pairwise interactions. 

The building blocks of a simplicial complex are the simplices. A {\em $n$-dimensional simplex} $\alpha$ (or $n$-simplex) is formed by a set of  $n+1$ nodes
\bea
\alpha=[v_0,v_1,\ldots, v_n],
\eea
with an assigned orientation.

Based on this definition, a $0$-simplex is a node, a $1$-simplex is a link, a $2$-simplex is a triangle, and so on. The $n'$-dimensional {\em faces} of a $n$-simplex $\alpha$ are defined as the simplices formed by a proper subset of the nodes in $\alpha$. Finally, a {\em simplicial complex} is a set of simplices closed under the inclusion of the faces of each simplex. The {\em dimension} $d$ of a simplicial complex is the largest dimension of its simplices. The simplices of a simplicial complex that are not faces of any other simplex are called {\em facets}.

Here and in the remainder of this work we indicate with $N_{[n]}$ the number of simplices of dimension $n$ present in a simplicial complex. Therefore, $N_{[0]}, N_{[1]}$ and $N_{[2]}$ indicate, respectively, the total number of nodes, links and triangles present in the simplicial complex.

As an example, a higher-order collaboration network can be described by a simplicial complex where one considers all the teams of co-authors of at least one paper as simplices, and includes the corresponding simplex and all its faces in the simplicial complex \cite{patania2017shape,carstens2013persistent}. Therefore, given an unweighted simplicial complex constructed in this way the collaboration network cannot be fully reconstructed, as only the facets will indicate for sure a higher-order collaboration.

Our aim is to show that weighted simplicial complexes are instead able to capture faithfully the higher-order collaboration data without any loss of information, provided that a proper choice of weights is made.

In a general framework, weighted simplicial complexes are enriched by {\em topological  weights} $w_{\alpha}>0$ associated to each simplex $\alpha$ of the simplicial complex.  The question that we want to address is: {\em how to best choose these topological weights without loosing the information present in higher-order network data?}

We assume to have as input data some {\em bare affinity weights} $\omega_{\alpha}\geq 0$ associated to each simplex of a simplicial complex. For instance, in the collaboration network that will be used for the present analysis, the bare affinity depend on the number of papers co-authored by the team represented by the generic simplex $\alpha$. Therefore, we can have a triangle $[1,2,3]$ with a positive bare weight $\omega_{[1,2,3]}>0$ indicating the existence of at least one paper written by the corresponding three authors. At the same time, we might also have one of more of its faces with null bare weights; for instance, we could have $\omega_{[1,2]}=0$, indicating that there are no two-author papers written by authors $1$ and $2$.
Starting from collaboration data, we propose a way to derive the topological weights $w_{\alpha}$, which are positive for every simplex of the simplicial complex. In the above example, for instance, we would have $w_{[1,2,3]}>0$, as well as $w_{[1,2]}>0$.
Given a $d$-dimensional simplicial complex, the proposed choice of topological weights  associated to the simplices $\alpha'$ of dimension $d$ is equal to the bare affinity weights 
\bea
w_{\alpha'}=\omega_{\alpha'}.
\label{wd}
\eea
However the topological weights of  simplices $\alpha'$ of dimension $n'=n_{\alpha'}<d$ are defined iteratively as the sum of the topological weights of the $n'+1$ dimensional simplices $\alpha$ incident to it plus the bare affinity weights $\omega_{\alpha'}$, i.e.
\bea
w_{\alpha'}=\sum_{\alpha\supset \alpha'}w_{\alpha}\tilde{\delta}({n_{\alpha},n_{\alpha'}+1})+\omega_{\alpha'},
\label{weigh:1}
\eea
where here $\tilde{\delta}(x,y)$ indicates the Kronecker delta, i.e. $\hat{\delta}(x,y)=1$ if $x=y$ and $\hat{\delta}(x,y)=0$ otherwise.
For example, in a simplicial complex of dimension $d=2$ formed by nodes, links and triangles, the topological weights associated to the triangles are the bare affinity weights, while the topological weights associated to the links are the sum of all the weights of the triangles incident to them, plus their bare affinity weights. Similarly, the topological weight of the nodes will be the sum of the topological weights of their incident links plus their bare affinity weight.

It is easy to check that since Eq.(\ref{wd}) and Eq.(\ref{weigh:1}) are linear, they are invertible. Therefore, with this choice of topological weights, it is always possible to reconstruct the bare affinity weights and have a faithful representation of the data, also if the data includes a set of higher-order interactions that is not closed under the inclusion of the subset of their nodes as in general collaboration data.

The topological weights of a simplicial complex can eventually evolve and fluctuate in time and, in that case, they are properly called {\em topological signals}, whose dynamics has recently attracted large attention \cite{bianconi2021higher,millan2020explosive,barbarossa2020topological,millan2021local,torres2020simplicial,calmon2021topological,bianconi2021topological,arnaudon2021connecting,deville2021consensus,ziegler2022balanced,reitz2020higher,schaub2020random,schaub2021signal,sardellitti2021topological,ghorbanchian2021higher}. In this paper, however, we will consider only topological weights constituted by single snapshots of topological signals, or by topological signals that are constant in time.

\section{Higher-order spectrum of weighted simplicial complexes}

Weighted simplicial complexes are able not only to faithfully represent higher-order network data without any loss of information, but allow also the investigation  of their higher-order spectrum thus revealing important properties of higher-order diffusion \cite{torres2020simplicial,reitz2020higher,ziegler2022balanced}. In this section we introduce the key algebraic topology background to study the higher-order spectrum of weighted simplicial complexes, which constitutes a fundamental pathway to relate higher-order structure to higher-order dynamics. 
Interesting background literature for this section include the References \cite{bianconi2021higher,josthorak2013spectra,zhang2022higher,lim2020hodge,grady2010discrete,hatcher2005algebraic}.

\subsection{Chains and Co-chains}

An $n$-chain is an element of the free abelian group  $C_{n}$ with basis the $n$ simplices of the simplicial complex defined with respect to the field $\mathbb{Z}$. Therefore, any $n$-chain $\sigma\in C_n$ can be written as a linear combination of $n$-simplicies with integer coefficients, i.e. 
\bea
\sigma=\sum_{\alpha}\sigma_{\alpha}\alpha,\eea
 with $\sigma_{\alpha}\in \mathbb{Z}$.
The boundary operator
$\partial_n:C_n\to C_{n-1}$ is a linear operator that maps $n$-chains to $(n-1)$-chains, and it is completely defined by its action on each $n$-simplex as follows:
\bea
\partial_n [v_0,\dots,v_n]=\sum_{p=0}^n (-1)^p [v_0,\ldots,\hat{v}_{p}\ldots, v_n],
\label{boundary}
\eea
where the notation $\hat{v}_p$ denotes the fact that vertex $v_p$ is missing from the simplex $[v_0,\ldots\hat{v}_{p}\ldots, v_n]$.
From Eq. (\ref{boundary}) it is clear that the boundary of a $n$-simplex is a $(n-1)$-chain formed by the $(n-1)$-dimensional simplices at its boundary, and oriented in the same way as the $n$-simplex.
One of the major topological properties of the boundary operator is that ``the boundary of the boundary is null" which translates into the following algebraic condition: 
\bea\label{pp}
\partial_{n}\partial_{n+1}=0.
\eea
Given a basis for the simplices of a simplicial complex, the boundary operator $\partial_n$ is represented by  incident matrices
$B_{[n]}$ which are  $N_{n-1}\times N_{n}$ rectangular matrices of elements 
\bea
B_n(\alpha',\alpha)=(-1)^p,
\eea
where $\alpha'$ and $\alpha$ are the simplicies 
\bea
\alpha'&=&[v_0,v_1,\ldots, \hat{v}_{p},\ldots, v_n],\\
\alpha&=&[v_0,v_1,\dots,v_n].
\eea
Since the boundary operator satisfies Eq.(\ref{pp}), we have 
\bea
B_{[n]}B_{[n+1]}=0,
\eea
for every $n>0$.

An $n$-cochain $f$ in the cochain group $C^{n}$ is an homeomorphism between the $n$-chains $C_n$ and the set of real numbers $\mathbb{R}$.
Given a $n$ chain $\sigma=\sum_{\alpha\in Q_{n}}\sigma_{\alpha}\alpha$ we have that  
\bea
f(\sigma)=\sum_{\alpha\in Q_{n}}\sigma_{\alpha}f(\alpha).
\eea
It follows that a $n$-cochain $f$ is uniquely determined by the vector ${\bf f}$ of elements given by $f_{\alpha}=f(\alpha)$.
The coboundary operator $\delta_n$ is a linear operator mapping  $n$ co-chains $f$, (i.e. linear functions defined on $n$-simplices) to $n+1$ cochains (i.e. linear functions defined on $n+1$ simplices). In particular, the $n$-coboundary operator 
$\delta_n:C^{n}\to C^{n+1}$ is defined by
\bea
\hspace{-6mm}(\delta_n f)[v_0,\dots,v_{n+1}]=a_n\sum_{p=0}^n (-1)^p f([v_0,\ldots, \hat{v}_{p},\ldots, v_{n+1}]),
\eea
where here we have introduced the constant $a_n\in \mathbb{R}^{+}$  that depends only on $n$ for later convenience. Typically, $a_n$ is taken to be one, namely $a_n=1$, but in the present setting $a_n$ can be assigned a value equal to any real positive constant.
For $a_n=1$ the coboundary operator $\delta_n$ is the dual of the boundary operator $\partial_{n+1} $, and for any value of $a_n$ it satisfies
\bea
(\delta_n f)[v_0,\dots,v_{n+1}]=a_n f(\partial_{n+1} [v_0,\dots,v_{n+1}]).
\eea
The topological properties that the ``boundary of the boundary is null" stated in Eq. (\ref{pp}) implies the following analogous algebraic property of the co-boundary operator:
\bea\label{dd}
\delta_{n+1}\delta_{n}=0.
\eea
Given a basis for the simplices of a simplicial complex, the coboundary operator $\delta_{n-1}$ is represented by the matrix
$\hat{B}_{[n]}$, which is a $N_{n}\times N_{n-1}$ rectangular matrix of elements 
\bea
\bar{B}_n(\alpha,\alpha')=a_n(-1)^p,
\eea
where $\alpha$ and $\alpha'$ are the simplices
\bea
\alpha'&=&[v_0,v_1,\ldots, v_{p-1}v_{p+1},\ldots, v_{n}],\\
\alpha&=&[v_0,v_1,\dots,v_{n}].
\eea
Therefore, we have  that $\bar{B}_{[n]}$ is simply related to the transposed of $B_{[n]}$, i.e.
\bea
\bar{B}_{[n]}=a_nB_{[n]}^{\top}.
\eea
Since the coboundary operator satisfies Eq.$(\ref{dd})$ we obtain 
\bea 
\bar{B}_{[n+1]}\bar{B}_{[n]}=0,
\eea
for every $n>0$.
\subsection{The coboundary operator and its dual}
We introduce now a non trivial ``metric" induced by the affinity weights $w_{\alpha}>0$ of the simplices $\alpha$ of the simplicial complex. Using a similar notation used in Grady and Polimeni \cite{grady2010discrete} we define the matrices $G_{[n]}^{-1}$ as the diagonal $N_{[n]}\times N_{[n]}$ matrices having as diagonal elements the topological weights of the $n$-simplices, i.e.
\bea \label{Gm0}
G_{[n]}^{-1}(\alpha,\alpha)=w_{\alpha}.
\eea
The matrices $G_{[n]}^{-1}$ are used to define a $L^{2}$ norm between $n$-cochains. In particular, indicating with $f^{(1)}$ and $f^{(2)}$ two $n$-dimensional cochain, the $L^2$ norm between the two cochains as
\bea
\langle f^{(1)},f^{(2)}\rangle=\sum_{\alpha\in Q_{n}}w_{\alpha}f^{(1)}_{\alpha}f^{(2)}_{\alpha}=({\bf f}^{(1)})^{\top}G_{[n]}^{-1} {\bf f}^{(2)}.
\eea
Based on this norm, the definition of $\bar{B}_{[n]}^*$ as the adjoint operator of $\bar{B}_{[n]}$ is derived. Formally, for any $n$-cochain $f$ and any $(n+1)$ cochain g, the adjoint operator $\bar{B}_{[n]}^*$ satisfies
\bea
\langle g,\bar{B}_{[n]}f\rangle=\langle \bar{B}_{[n]}^*g,f\rangle.
\label{uno}
\eea
From this definition we deduce the explicit expression of  $\bar{B}_{[n]}^*$ in terms of the coboundary operator $\bar{B}_{[n]}$ and the matrices $G_{[n]}$. Indeed, given Eq. (\ref{uno}) we obtain
\bea
{\bf g}^{\top}G_{[n]}^{-1} \bar{B}_{[n]} {\bf f}={\bf g}^{\top} (\bar{B}_{[n]}^*)^{\top} G_{[n-1]}^{-1} {\bf f}.
\eea
Since this expression should hold for any arbitrary $f$ and $g$, we obtain
\bea
G_{[n]}^{-1} \bar{B}_{[n]}=(\bar{B}_{[n]}^*)^{\top} G_{[n-1]}^{-1},
\eea
from which we get the explicit expression of the adjoint of the coboundary operator given by 
\bea
\bar{B}_{[n]}^*=G_{[n-1]}\bar{B}_{[n]}^{\top}G_{[n]}^{-1}.
\eea
This operator \cite{josthorak2013spectra} is sometimes referred to in the literature as the ``weighted boundary operator" \cite{kelinbio2020weighted,kelin018weighted}. In fact,  if the metric matrices are trivial, i.e, $G_{[n]}={I}$, the above expression reduces to the boundary operator multiplied by $a_n$, i.e.
\bea
\bar{B}_{[n]}^*=\bar{B}_{[n]}^{\top}=a_nB_{[n]}.
\eea

\subsection{Higher-order Weighted Laplacians and Hodge decomposition}
The graph Laplacian is a well known operator that describes diffusion from nodes to nodes through links in a network. Higher-order Laplacians $L_{[n]}$, also called Hodge Laplacians \cite{josthorak2013spectra,lim2020hodge,schaub2020random}, generalize the notion of graph Laplacian to describe higher-order diffusion. For instance, if we consider diffusion from links to links ($n=1$), the Hodge Laplacian can describe diffusion through nodes or through triangles.
On a $d$-dimensional simplicial complex  the higher-order Laplacian (or Hodge Laplacian) $L_{[n]}$ is defined for each $n=0,\ldots,d$ as
\bea\label{eq::holap}
    L_{[0]}&=&L_{[0]}^{up}\nonumber, \\
    L_{[n]}&=&L_{[n]}^{down}+L_{[n]}^{up}\ \mbox{for }\  n>0,
\eea
where $L_{[n]}^{down}$  and $L_{[n]}^{up}$ describe diffusion from $n$-simplices to $n$-simplices through $(n-1)$ simplices and  $(n+1)$ simplices, respectively. They are formally defined as
\bea
L_{[n]}^{down}&=&\bar{B}_{n}\bar{B}_{n}^{*},\nonumber \\
L_{[n]}^{up}&=&\bar{B}_{n+1}^*\bar{B}_{n+1}.
\label{bb1}
\eea
From the definition of the higher-order Laplacians (Eq.(\ref{eq::holap})) and from Eqs. (\ref{bb1}) it follows immediately that not only $L_{[n]}^{up}$ and $L_{[n]}^{down}$ commute, but they also obey the additional stronger  property that 
\bea
L_{[n]}^{up}L_{[n]}^{down}&=&0,\nonumber \\
L_{[n]}^{down}L_{[n]}^{up}&=&0.
\eea
This property implies that 
\bea
\mbox{ker}L_{[n]}^{up}&\supseteq&\mbox{im}L_{[n]}^{down},\nonumber \\
\mbox{ker}L_{[n]}^{down}&\supseteq&\mbox{im}L_{[n]}^{up}.
\eea
Therefore any eigenvector of $L_{[n]}$ corresponding to a non-zero eigenvalue $\lambda>0$ is either a non-zero eigenvector of $L_{[n]}^{up}$ or a non-zero eigenvector of $L_{[n]}^{down}$ with the same eigenvalue $\lambda$.
This implies that the set of cochains $C^n$ obeys the Hodge decomposition, as we have 
\bea
C^n=\mbox{im}(\bar{B}_{[n]})\oplus\mbox{ker}(L_{[n]})\oplus \mbox{im}(\bar{B}_{[n+1]}^*).
\eea
For example, a $n=1$ signal can be decomposed into a gradient flow, an harmonic flow and a solenoidal flow.

\subsection{The Weighted Dirac operator}

The topological Dirac operator \cite{bianconi2021topological,lloyd2016quantum,ameneyro2022quantum,knill2013dirac} is an important topological operator that can be interpreted as the ``square root" of the Laplacian and can be used to treat simultaneously  topological signals of different dimensions.
On a weighted simplicial complex we define the weighted topological Dirac operators as the linear operator acting on the direct sum of all the $n$-cochains defined in the system, i.e. acting in the linear space $\oplus_{n=1}^{d}C^n$ and having as matrix representation the $M\times M$ matrix where $M=\sum_{n=1}^d N_{[n]}$ of elements 
\bea
D_{\alpha,\alpha'}=\left\{
\begin{array}{ccc}
\bar{B}^{*}_{[n]}({\alpha,\alpha'}) 
& \mbox{if} & n_{\alpha'}=n=n_{\alpha}+1,\\
\bar{B}_{[n]}({\alpha,\alpha'})& \mbox{if} & n_{\alpha}=n=n_{\alpha'}+1.
\end{array}\right.
\eea
In the case of a simplicial complex of dimension $d=2$ including nodes, links and triangles, the weighted topological Dirac operator has dimension $M\times M$, with $M=N_{[0]}+N_{[1]}+N_{[2]}$, and a block matrix structure of the form
\bea
D=\left(\begin{array}{ccc}0& \bar{B}_{[1]}^* &0\\
\bar{B}_{[1]}& 0 & \bar{B}_{[2]}^*\\
0& \bar{B}_{[2]} & 0\end{array}\right).
\eea
From the definition of the weighted topological Dirac operator one concludes that  
the square of the Dirac operator is the direct sum of the higher-order Laplacians
\bea
D^2=L_{[1]}\oplus L_{[2]}\oplus L_{[3]}\oplus \ldots L_{[n]}.
\eea
This implies for a simplicial complex of dimension $d=2$ that 
\bea
D^2=\left(\begin{array}{ccc} {L}_{[0]} &0 &0\\
0&{L}_{[1]}& 0  \\
0& 0 & {L}_{[2]} \end{array}\right).
\eea
Moreover it follows that the weighted topological Dirac operator is self-adjoint, i.e. 
\bea
D^*=D.
\eea
The eigenvalues $\lambda$ of the weighted topological operators are in absolute value equal to the singular values of the coboundary operators of the simplicial complex under consideration.

\subsection{Normalized Higher-order Laplacians }
Defining normalized Laplacians is a central theme in spectral graph theory \cite{chung1997spectral}.
For graphs, the Hodge Laplacian $L_{[0]}=\bar{B}_{[0]}^*\bar{B}_{[0]}$ depends on the choice of suitable  metric matrices. For the choice of the metric matrices $G_{[0]}=I$ and $G_{[1]}=I$, and for $a_1=1$, we obtain the un-normalized (or combinatorial) Laplacian 
\bea
L_{[0]}={K}_{[0]}-A,
\eea
where $K_{[0]}$ is the diagonal matrix having the degree of the nodes  as diagonal elements and where $A$ is the weighted adjacency matrix of the network.
The normalized (weighted) Laplacian can instead be obtained by considering the weights of the links $w_{ij}$ as arbitrary (positive) values, the weight of the nodes as given by the node strength, i.e. the sum of the weights of the incident links, and imposing also that the metric matrices given by Eq. (\ref{Gm0}), i.e.
\bea
G_{[1]}^{-1}([i,j],[i,j])&=&w_{ij},\nonumber \\
G_{[0]}^{-1}([i],[i])&=&w_i=\sum_{j=1}^{N_{[0]}}w_{ij}.
\label{m_norm}
\eea 
This choice of the metric matrices, together with the choice $a_1=1$, implies that 
\bea
L_{[0]}=I-{K_{[0]}}^{-1}A,
\eea
where now $K_0^{-1}=G_{[0]}$ is the diagonal matrix having the inverse of the strength of the nodes as diagonal elements.
This normalized Laplacian  is well known to have a real bounded spectrum with eigenvalues $0\leq \lambda\leq 2$ (see discussion in \cite{chung1997spectral}).
This implies that by considering the metric matrices given by Eq. (\ref{m_norm}), but taking $a_1=\sqrt{2}$, we obtain a Laplacian matrix whose eigenvalues are non-negative and  not larger than one, with 
\bea
L_{[0]}=\frac{1}{2}\left[I-{K_{[0]}}^{-1}A\right],
\eea
One important question is whether  we can follow similar arguments to propose a normalized version of the higher-order Laplacian.
In general, proposing a normalized higher-order Laplacian is an important problem in graph theory. However, this issue has only been addressed in a few papers until now (see for instance
\cite{josthorak2013spectra,schaub2020random}).
Here we show that with our choice of topological weights given by Eq. (\ref{wd}) and Eq. (\ref{weigh:1}), by using metric matrices whose diagonal elements are given by Eq.(\ref{Gm0}), the  higher-order Hodge Laplacians are automatically normalized provided that we choose 
\bea
a_n=(n+1)^{1/2},
\eea
for any $n$ up to the order of the simplicial complex.
Under these hypotheses, it can be shown that the eigenvalues of the Hodge Laplacians  are always in the interval $[0,1]$ (see Appendix \ref{Ap:normalization}). 

Note that the choice of topological weights proposed in this work, i.e. Eq.(\ref{wd}) and Eq.(\ref{weigh:1}), naturally generalizes to the higher order case the choice of weights that is usually adopted for normalizing the graph Laplacian that, as discussed above, assigns to nodes the sum of the weights of the incident links.

\section{Higher-order spectral entropy of simplicial complexes}
\label{sec:entropy}
\subsection{Definition of higher-order spectral entropy}
In order to characterize the information content encoded in the spectrum of networks and generalized network structures, the spectral entropy, also called Von Neumann entropy of networks, has been introduced.
\cite{anand2009entropy,anand2011shannon,de2015structural,
de2016spectral,ghavasieh2020statistical,ghavasieh2022statistical}.
The spectral entropy of a network is defined as the quantum mechanics von Neumann entropy \cite{nielsen2002quantum}, where the density operator is taken to be a semi-definite positive operator associated to the network and having normalized trace. Therefore, typical choices for the density operator are taken to be functions of the graph Laplacian. 
Here we define the higher-order spectral entropy of weighted simplicial complexes and use this quantity, together with the associated higher-order relative entropy, in order to evaluate the information content of the higher-order spectrum of weighted simplicial complexes.

Given a simplicial complex of order $d$, we define the higher-order spectral density  $\rho_n$   as:
\bea\label{eq::netdensity}
    \rho_{n} = \frac{e^{-\beta L_{[n]}}}{Z_{n}},
\eea
where $Z_{n}= Tr[e^{-\beta L_{[n]}}]$. Note that for $n=0$ this definition reduces to the spectral density of networks proposed in \cite{de2016spectral}. 

The spectral  entropy of order $n$, also called $n$-order von Neumann entropy,  is then defined as:
\bea\label{eq::entropy}
    S_n = -Tr[\rho_{n}\ln{\rho_n}].
\eea
Note that here and in the following we choose for convenience to use the natural logarithm, as it is common practice in machine learning and statistical mechanics \cite{goodfellow2016deep,kardar2007statistical}. 
As usual for the von Neumann entropy it is straightforward to derive the following compact form of $S_n$:
\bea\label{eq::finalformvn}
    S_n=\beta\Avg{\lambda}_n+\ln Z_n,
\eea
where 
\bea\label{lambdamean}
    \Avg{\lambda}_n=\frac{\sum_{i}e^{-\beta\lambda_i(L_{[n]})}\lambda_i(L_{[n]})}{Z_n},
\eea
and $\lambda_i(L_{[n]})$ denotes the generic $i-$th eigenvalue of $L_{[n]}$. It is useful to recall that based on the Hodge decomposition, if $n>0$ the eigenvalues of the higher order Laplacian $L_{[n]}$ can be partitioned into:
\begin{itemize}
    \item[-]Zero eigenvalues (\textit{harmonic}), denoted as $\left\{\lambda_h\right\}$;
    \item[-]Nonzero eigenvalues of $L_{[n]}^{up}$, denoted as $\left\{\lambda_{u}\right\}$;
    \item[-]Nonzero eigenvalues of $L_{[n]}^{down},$ denoted as $\left\{\lambda_{d}\right\}$
\end{itemize}
In the case $n=0$, instead, the eigenvalues are either zero (harmonic eigenvalues) or non-zero eigenvalues of $L_{[0]}^{up}=L_{[0]}$.

\subsection{Spectral density  and return time distribution}\label{sec::interpret_beta}

The adoption of Eq. (\ref{eq::netdensity}) as a network density can be interpreted in terms of a higher-order diffusion process, as we will discuss in this paragraph. From this observation we will be able to derive a higher-order relation between the entropy, the specific heat and the temporal scales of higher-diffusion processes which represent the higher-order version of the analogous relations on networks \cite{villegas2022path}. Higher-order diffusion \cite{torres2020simplicial,reitz2020higher,ziegler2022balanced,mulder2018network,millan2021local} describes diffusion from $n$-simplices to $n$-simplices going either though $n-1$ simplices or $n+1$ simplices. The diffusion corresponding dynamics is dictated  by the  Laplacian $L_{[n]}$ as described by the following system of differential equations \cite{torres2020simplicial}:
\begin{equation}\label{eq::diffusion}
\dot{{\bf X}}(t)=- L_{[n]} {\bf X}(t),
\end{equation}
where $X(t)$ is a column vector representing a $n$-cochain, and can be seen as a function describing the distribution of information on the $n$-dimensional simplices of a simplicial complex at time $t$. By solving the above equation and using the spectral decomposition of the Hodge Laplacian, we can express ${\bf X}(t)$ as 
\begin{equation}
    {\bf X}(t)=\sum_{\lambda}c_\lambda(t) {\bf u}_{\lambda},
\end{equation}
where ${\bf u}_{\lambda}$ is the eigenvector associated to the eigenvalue $\lambda$ of $L_{[n]}$, and every $c_\lambda$ follows the temporal evolution
\begin{equation}\label{eq::diffusion_solved}
    c_\lambda(t)=e^{-\lambda t}c_\lambda(0).
\end{equation}

Let us suppose that a random walker starts a walk in the network at a simplex $\alpha_0$. Then, we will have
\begin{equation}\label{eq::rettime}
    X_\alpha(t) = \sum_{\lambda} c_\lambda(0)e^{-\lambda t}u_{\alpha}^{\lambda}.
\end{equation}
 We observe that at time $t=0$ we have:
\begin{equation}\label{eq::tzero1}
   X_\alpha(0)=\delta_{\alpha,\alpha_0}=\sum_{\lambda} c_\lambda(0)u_\alpha^{\lambda} 
\end{equation}

From Eq. (\ref{eq::tzero1}) we derive that $c_{\lambda}(0)=u_{\alpha_0}^{\lambda}$; thus, $X_{\alpha}(t)$ can be expressed as
\begin{equation}\label{eq::tzero}
   X_\alpha(t)=\sum_{\lambda} e^{-\lambda t}u_{\alpha_0}^{\lambda}u_{\alpha}^{\lambda}.
\end{equation}

Finally, the return time distribution on simplex $\alpha_0$ is obtained as: 
\begin{equation}\label{eq::return_time_final}
    X_{\alpha_0}(t)=\sum_{\lambda} e^{-\lambda t}u_{\alpha_0}^{\lambda}u_{\alpha_0}^{\lambda}.
\end{equation}

Taking the average over all the possible $n$-dimensional simplices and using the normalization of the eigenvectors of $L_{[n]}$ we obtain that return time probability $p(t)$ is given by 
\begin{equation}\label{eq::avg_rettime}
    p(t)=\frac{1}{N}\sum_{\lambda} e^{-\lambda t}.
\end{equation}

It follows that, classically, it is possible to interpret the  density operator $\rho_n$ in terms of the average return time distribution. Indeed, since $Z_n=\sum_{\lambda}e^{-\beta\lambda}$, if we interpret $\beta$ as time $t$, $Z_n$ can be seen as the return time distribution associated to the higher-order diffusion of order $n$. 
By consequence, the density $\rho_n$ introduced in Eq. (\ref{eq::netdensity}) tells how much each eigenvalue contributes to the return time probability $p(t)$ for $t=\beta$, and the spectral entropy tells how many eigenvalues contribute significantly to the return time distribution.
In the analysis of the spectral entropy of a  network it has been recently proposed \cite{villegas2022path} to monitor the specific heat whose local minima and maxima capture the characteristic scale of diffusion on nodes and links.
Here we propose to use the higher-order specific heat $C_n$ given by the derivative of the higher-order spectral entropy 
\bea
C_n=\frac{\partial S_n}{\partial \beta},
\eea
to monitor and to  characterize  the typical temporal scales of  higher-order diffusion on simplicial complexes.

\subsection{Relative entropy}\label{sec::rel_entropy}
In quantum information theory \cite{nielsen2002quantum} the  Von Neumann relative entropy (or quantum Kullback-Leibler divergence) between two densities operators $\rho$ and $\sigma$ acting over the same space is defined as 
\begin{equation}\label{eq::re}
    {KL}(\rho\|\sigma) = Tr[\rho(\ln\rho-\ln\sigma)].
\end{equation}
Note that analogously to the classical Kullback-Leibler divergence, the quantum relative  entropy ${KL}(\rho\|\sigma)$ is not symmetric.
When associating density operators to networks, the condition that $\rho$ and $\sigma$ should act on the same space imposes that the number of nodes of the two considered networks must be the same.
In particular, in Ref. \cite{de2016spectral} the spectral relative entropy of networks has been studied to perform a pairwise comparison of the layers of a multiplex network.
In our setting, as we will show in the remainder of this analysis, the quantum relative entropy ${KL}(\rho\|\sigma)$ is suitable to perform a comparison between density matrices constructed on higher-order Laplacians starting from the weighted or the unweighted version of the same dataset (i.e. a dataset in which the bare affinity weights can be chosen to be heterogeneous, or instead to be homogeneous and all equal to one).

When studying the spectral properties of simplicial complexes, it is desirable to compare also the density operators associated to Hodge Laplacians of different orders, with the aim of revealing differences in the diffusion processes occurring at different dimensions. 
However, the generalization of Eq. (\ref{eq::re}) is not straightforward, as $\rho_n$ and $\rho_{n+1}$ act on spaces of different dimension.

To overcome the dimensional incompatibility, we propose to  consider the  projected density operator $\hat{\rho}_{n+1,n}$ given by 
\begin{equation}\label{def::projection}
    \hat{\rho}_{n+1,n}=\bar{B}_{n+1}^*\frac{e^{-\beta L_{[n+1]}} }{\hat{Z}_n}\bar{B}_{n+1},
\end{equation}
where  $\hat{Z_n}$ is the normalization constant given by 
 \begin{equation}
    \hat{Z_n}= \mbox{Tr}[\bar{B}_{n+1}^* e^{-\beta L_{[n+1]}} \bar{B}_{n+1}].
    \end{equation}
Using the projected density operator $\hat{\rho}_{n+1,n}$ we can use the quantum  relative entropy  to compare the densities operators associated to  different orders:
\begin{equation}\label{eq::relentropy_withproj}
{KL}(\hat{\rho}_{n+1,n}||\rho_{n}) = \mbox{Tr}[\hat{\rho}_{n+1,n}(\ln{\hat{\rho}_{n+1,n}}-\ln{\rho_{n}})].
\end{equation}
This relative entropy can be also written as \bea
{KL}(\hat{\rho}_{n+1,n}||\rho_{n})=J(\hat{\rho}_{n+1,n}||\rho_{n})-\hat{S}(\hat{\rho}_{n+1,n}),\eea where $\hat{S}(\hat{\rho}_{n+1,n})$ is the Von-Neumann entropy associated to the projected density operator $\hat{\rho}_{n+1,n}$ and $J(\hat{\rho}_{n+1,n}||\rho_{n})$ is the quantum cross-entropy, i.e.
\bea
\hat{S}(\hat{\rho}_{n+1,n})&=&-Tr[\hat{\rho}_{n+1,n}\ln\hat\rho_{n+1,n}],\nonumber \\
J(\hat{\rho}_{n+1,n}||\rho_{n})&=&-Tr[\hat{\rho}_{n+1,n}\ln\rho_{n}].
\eea
A straightforward calculation shows that we can express $\hat{S}(\hat{\rho}_{n+1,n})$ and $J(\hat{\rho}_{n+1,n}||\rho_{n})$ as 
\bea\label{eq::computation_projection}
    \hat{S}(\hat{\rho}_{n+1,n})&=&\beta \Avg{\lambda_{n,u}}_{\uparrow} -\Avg{\ln{\lambda_{n,u}}}_{\uparrow} + \ln{\hat{Z}_n}.\nonumber \\
    J(\hat{\rho}_{n+1,n}||\rho_{n})&=&\beta \Avg{\lambda_{n,u}}_{\uparrow}+\ln Z_{n},
\eea
in which we have used the notation 
\bea
    \Avg{\lambda_{n,u}}_{\uparrow}&=&\frac{\sum_{i=1}^{M}e^{-\beta\lambda_{n,u}^{(i)}}\left(\lambda_{n,u}^{(i)}\right)^2}{\hat{Z}_n},\nonumber \\
    \Avg{\ln{\lambda_{n,u}}}_{\uparrow} &=&\frac{\sum_{i=1}^{M} e^{-\beta\lambda_{n,u}^{(i)}}\lambda_{n,u}^{(i)}\ln{\left(\lambda_{n,u}^{(i)}\right)}}{\hat{Z}_n},  
\eea
where $M$ denotes the number of non-zero eigenvalues of the $L_{[n]}^{up}$ Laplacian.
Note that while $\hat{Z}_n$ only depends on the non-zero spectrum of $L_{[n]}^{up}$, $Z_n$ depends both on the non zero spectrum of $L_{[n]}^{up}$ and to the non zero spectrum of $L_{[n]}^{down}$.

\section{Application to higher-order collaboration networks}
\subsection{The topological weights of higher-order collaboration networks}
\label{subsec::weighted_data}

\begin{figure*}[t]
    \centering
     \includegraphics[width=0.8\textwidth]{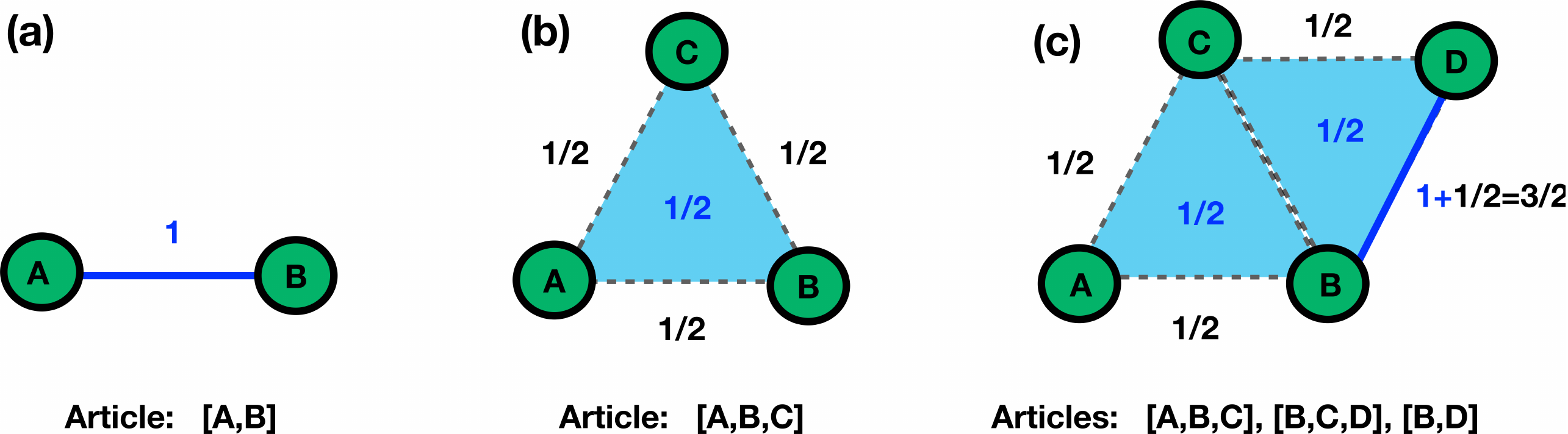}
     \caption{ Examples of bare affinity weights (indicated in blue) and topological weights (indicated in black) for higher-order collaboration networks captured by simplicial complexes of dimension $d=2$.
      Nodes, labeled with capital letters, represent authors; a blue link indicates a collaboration between two authors which coauthored a two-author paper; a filled triangle represents a collaboration among three authors leading to at least one three author paper. In panel (a) we consider the trivial case where a single article written by two coauthors is considered. Panel (b) represents the case of a single article written by three coauthors, while panel (c) is a more general situation where $3$ papers are considered, each of which with a variable number of authors. In each of the studied cases, it can be verified that the topological weight of a node given by the sum of the topological weights of the links incident to it, is given by the number of papers written by that node (author). From panel (c) it results that the topological weight of a link results in general (see case of the [BD] link) from the sum of the topological  weights of the triangles incident to it and the bare affinity weight.
     }
     \label{fig:coll}
\end{figure*}

Let $Q$ be the set of simplices formed by the teams of co-authors of the papers included in a collaboration network, where $p$ is the largest dimension of a simplex in $Q$.
First, let us work under the assumption that the higher-order collaboration network is encoded in a simplicial complex of dimension $d=p$. In this ideal hypothesis, the higher-order collaboration network is a simplicial complex $\mathcal{K}_p$ including all the simplices belonging to $Q$, together with all their faces.
We indicate with $m_n(\alpha)$ the number of papers written by the team of $n+1$ co-authors forming the simplex $\alpha$ of dimension $n_\alpha=n$, and we indicate with $\omega_{\alpha}$ a measure of the strength of the collaboration of team of co-authors  $\alpha\in Q$.
The aim is to define the correct definition of the bare affinity weights $\omega_{\alpha}$ such that the topological weight (given by Eq.(\ref{weigh:1})) associated to each node of the simplicial complex is equal to the number of papers written by the corresponding author.
With the definition of the topological weight given by Eq.(\ref{weigh:1}) it is easy to show that since each $n$-simplex includes $n$ faces of dimension $n-1$, if we want to enforce a topological weight of the nodes equal to the number of paper written by the corresponding author,  the strength of the collaboration encoded by the simplex $\alpha\in Q$ is uniquely defined as  
\bea
\omega_{\alpha}=\frac{m_n(\alpha)}{n_{\alpha}!}.
\label{weigh:2}
\eea
For instance the strength of a collaboration of a team formed by two authors is just equal to the number of two-author papers they have written together, while the strength of a collaboration of a team of three authors is equal to the number of three-authors papers they have wrote together divided by two, etc. (see Figure \ref{fig:coll}).

Note that both the recursive equations Eq.(\ref{weigh:1}) and Eq. (\ref{weigh:2}) are also perfectly defined  $n_{\alpha'}=0$ and for $n_{\alpha}=0$, thus allowing to consider also single author papers.

Since in higher-order collaboration data $p$ might be large, for computational reasons it might be convenient to encode the data in a simplicial complex $\mathcal{K}_d$ of dimension $d$ strictly smaller than $p$. In the special case in which $d=1$ the  collaboration network  $\mathcal{K}_1$ is obtained, in which each pair of nodes is connected if the linked authors have written a paper together. The collaboration network $\mathcal{K}_1$ retains only the nodes and links of $\mathcal{K}_p$ and is referred to as the network skeleton of $\mathcal{K}_p$. Similarly, in the case $d=2$, the simplicial complex $\mathcal{K}_2$ , will be considered, which is the $2$-skeleton of $\mathcal{K}_p$ retaining only the nodes, links and triangles present in  $\mathcal{K}_p$.

Our goal is to define a proper measure of the strength of the collaboration $\omega_{\alpha}$ associated to the simplices of $\mathcal{K}_d$, so that the topological weights defined recursively by Eq. (\ref{weigh:1}) attribute to each node a topological weight given by the number of the paper they have co-authored.

In this case we want to recover the normalization proposed in Ref. \cite{newman2001scientific} for collaboration networks, where the link weights are given by 
\bea
\omega_{\alpha}=\sum_{\alpha'\supset \alpha}\frac{m_{1,n'}(\alpha')}{n'}.
\label{old}
\eea
Here $m_{1,n}(\alpha')$ indicates how many papers a team $\alpha'$ of $(n+1)$ co-authors, including the two authors of the link $\alpha$ have written together.

In the general case in which the higher-order collaboration network is the $d$ skeleton  $\mathcal{K}_d$, we can generalize Eq.(\ref{old}) in the following way.
Let $m_{n,n'}(\alpha')$ denote the number of papers written by the team represented by the $n'$-simplex $\alpha'$ including the $n+1$ authors of the group of co-authors $\alpha$, with $n\leq n'$. 
A simple combinatorial calculation provides  the strength of collaboration of the simplex $\alpha\in \mathcal{K}_d$ that ensures that the topological weights defined in Eq. (\ref{weigh:1}) are consistent with a topological weight of the nodes that is equal to the number of papers written by that author. In particular, for $n_{\alpha}<d$ Eq. (\ref{weigh:2}) applies, while  for $n_{\alpha}=d$ we get 
\bea\label{def_weights_definitiva}
    \omega_{\alpha} = \frac{1}{d!}\sum_{\alpha'\supseteq \alpha} \frac{m_{d,n'}(\alpha')\theta(n',d)}{{\left(\begin{array}{c}{n'}\\{d}\end{array}\right)}}, 
    \eea
where $\theta(x)=1$ if $x\geq 0$ and $\theta(x)=0$ if $x<0$

Figure \ref{fig:coll} is an example helping to visualize the result of the weight assignment. Starting from the left, the trivial case of a single paper written by two coauthors is depicted. Then, the case of a paper written by three authors, followed by a more general situation, are depicted. It is easily verified from case (c) that the bare affinity weight of a link can be obtained by subtracting the topological weight of the link by the sum of the topological weights of the triangles incident to it.

\subsection{Higher-order collaboration dataset}
In order to perform our analysis, we build our own collaboration dataset starting from a list of articles downloaded from Scopus (\url{https://www.scopus.com/} accessed on 2022-01-29). Specifically, we selected data and metadata of articles whose metadata contained the expression ``multilayer network'', referred to five years (2017-2021). Only articles published on scientific journals were selected, while papers published on conference acts were excluded, for a total of $686$ articles. The original and processed data are available and can be downloaded from \url{https://github.com/DedeBac/WeightedSimplicialComplexes.git}.
In view of studying collaboration between authors, articles with only one author were filtered out. This results into $2,169$ authors with at least one collaboration. 
We encode the higher-order collaboration data into a weighted  simplicial complex of dimension $d=2$ where the affinity weights are choosen accounding to Eq.(\ref{def_weights_definitiva}) for triangles (with $n=d$)  and Eq.(\ref{weigh:2}) nodes and links (with  $n<d$).
Overall, the simplicial complex consists of $2,169$ vertices, $5,296$ edges and $9,279$ triangles.

\begin{figure*}[t]
    \centering
    \includegraphics[width=\textwidth]{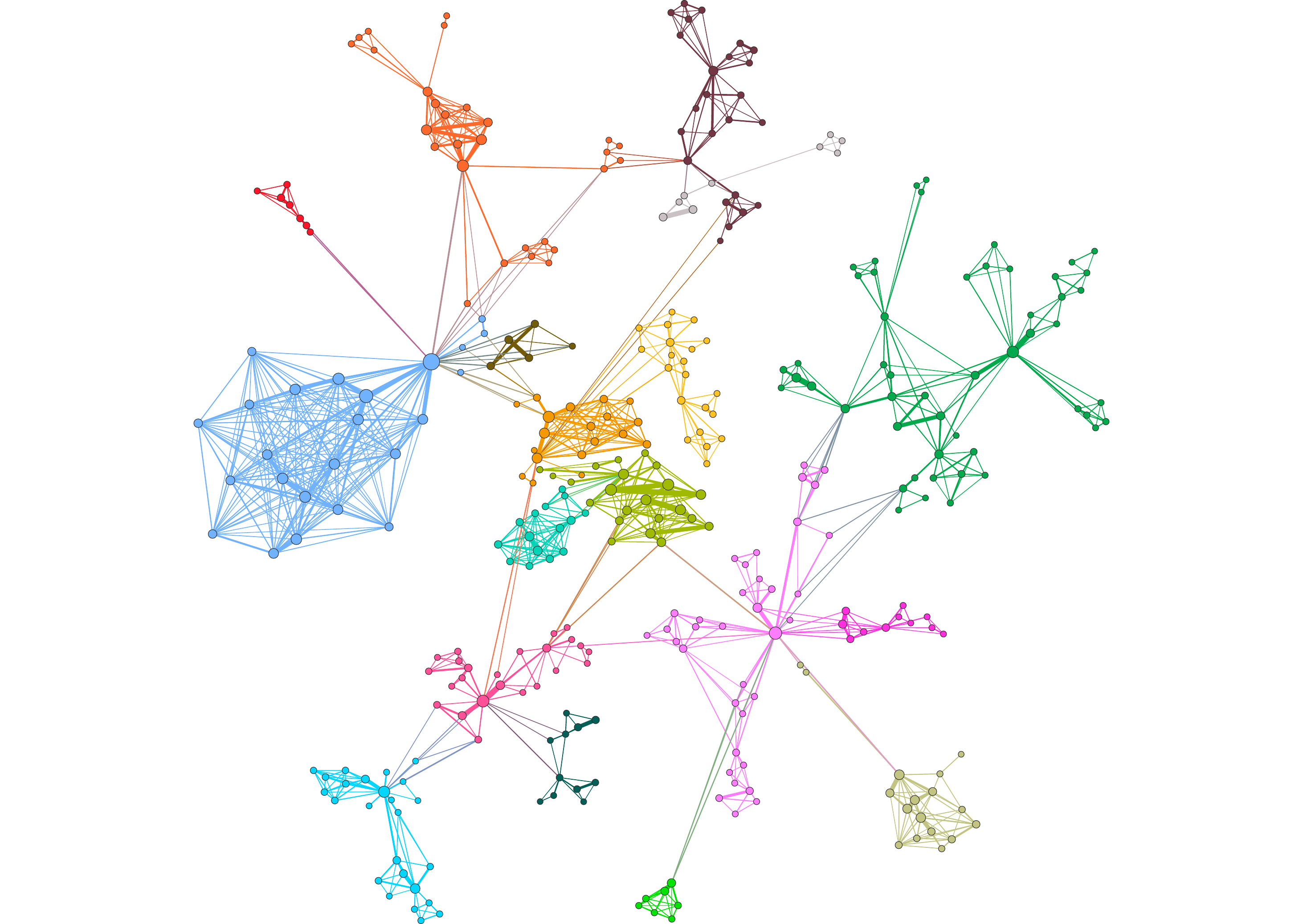}
    \caption{Network skeleton of weighted collaboration simplicial complex. Nodes and edges are coloured according to their modularity class. Size of nodes is proportional to their weighted degree. Edge thickness is proportional to the edge weight. The network is plotted with the Gephi software \cite{gephi} using the ForceAtlas2 visualization algorithm \cite{jacomy2014forceatlas2}. Different colours correspond to distinct communities computed using the Louvain algorithm \cite{blondel2008fast} by setting the resolution parameter to $1$ \cite{lambiotte2008laplacian}.The modularity score associated to the partition is $0.873$. Here for clarity we represent the network as unlabelled. For a  labelled version of the network see \url{https://github.com/DedeBac/WeightedSimplicialComplexes.git}.}
    \label{fig::weighted_collab}
\end{figure*}

As expected, the network skeleton of the simplicial complex is a disconnected graph, reflecting the fact that there might be isolated groups of collaborating authors. Since the study of the Von Neumann entropy can be interpreted in terms of information diffusion through the simplicial complex, we restrict our analysis to the simplicial complex originated by the largest connected component of the network skeleton.   
In  Figure \ref{fig::weighted_collab} we show the   network skeleton of the considered simplicial complex that has a rich community structure. The resulting simplicial complex has much smaller size than the original dataset, and consists of $356$ nodes, $1,172$ edges, and $2,614$ triangles. 

\subsection{Higher-order spectral entropy of the higher-order collaboration network}\label{subsec::exp_coauth}

\begin{figure*}[!ht]
    \centering
     \includegraphics[width=\textwidth]{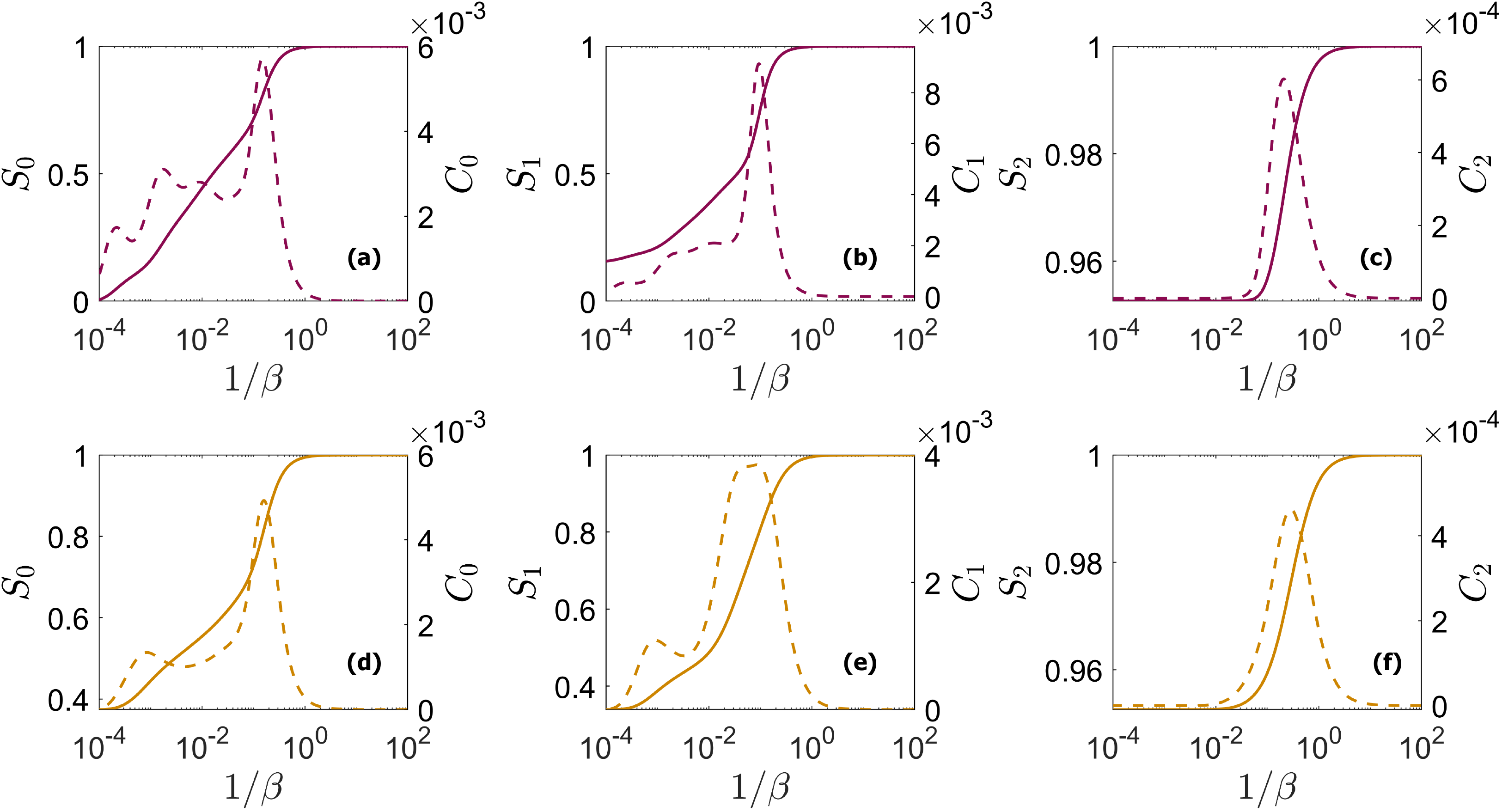}
     \caption{Von Neumann entropy of order 0 ((a), (c)), 1 ((b), (d)), and 2 ((e), (f)) of collaboration network as a function of $1/\beta$ ($\beta$ is the density parameter). Purple lines refer to the unweighted version of the network. Yellow lines refer to the weighted version. Dotted lines represent the derivative of the entropy with respect to the logarithm of $\beta$.}
     \label{fig::collab:VNentr}
\end{figure*}

\begin{figure*}[!ht]
    \centering
     \includegraphics[width=0.8\textwidth]{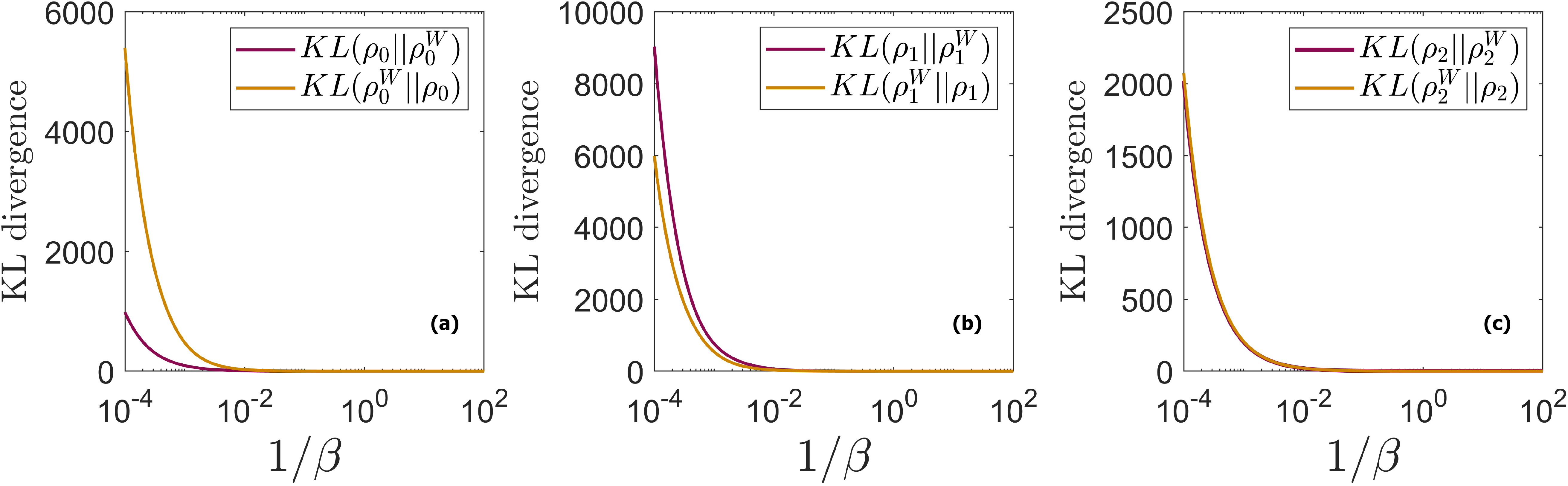}
     \caption{Relative entropy between weighted and unweighted simplicial complexes of collaboration data as a function of $1/\beta$. (a), (b), and (c) depict the relative entropy between the unweighted and weighted densities ($\rho_n$ and $\rho_n^{W}$, $n=0,1,2$) for the Hodge Laplacian of order $0, 1, 2$, respectively. Purple lines represent the quantity $KL(\rho_n||\rho_n^{W})$; yellow lines represent the quantity $KL(\rho_n^{W}||\rho_n)$.}
     \label{fig::rel_entWvsUW}
\end{figure*}

\begin{figure*}[!ht]
\centering
    \includegraphics[width=0.7\textwidth]{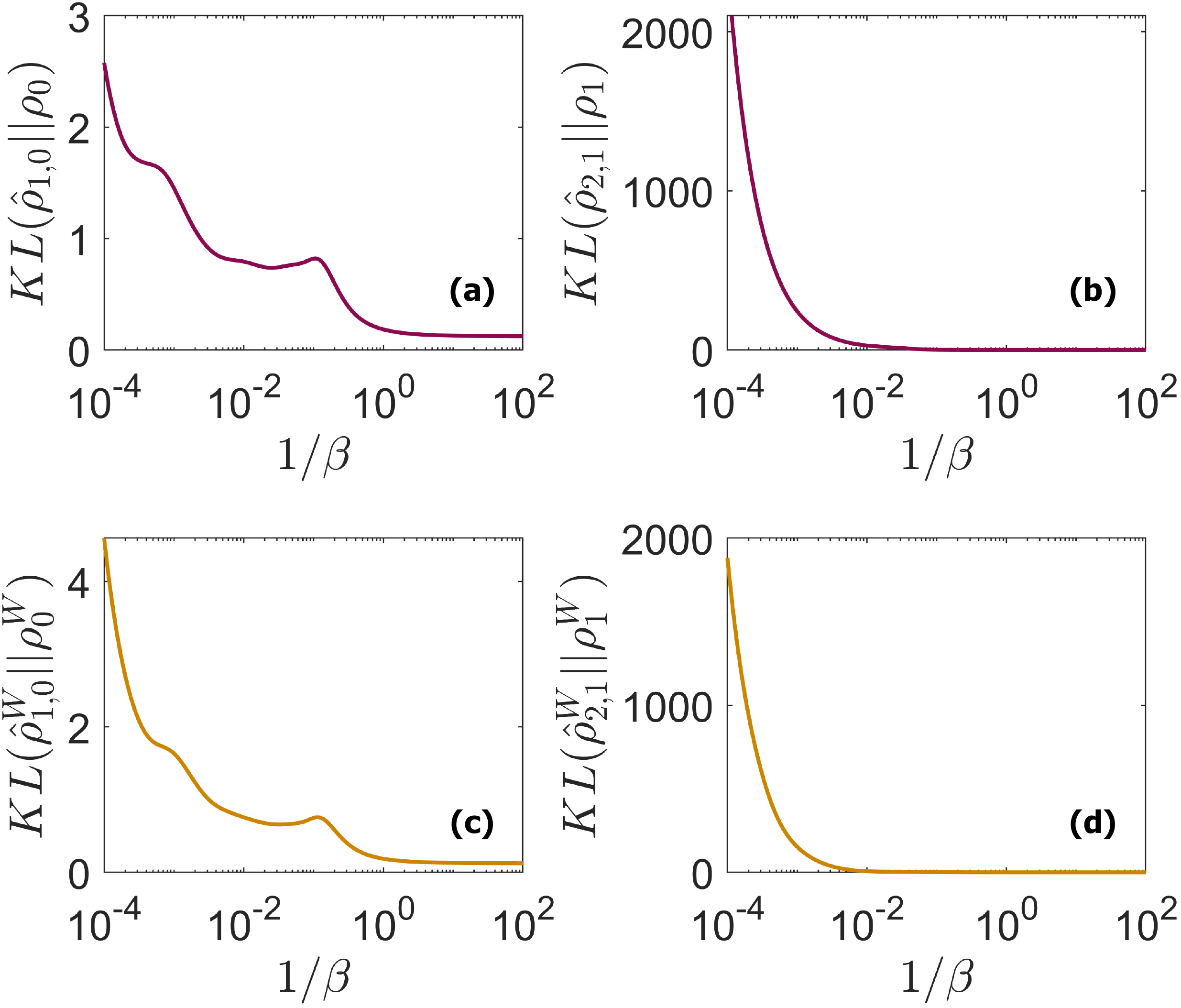}
     \caption{Each plot depicts the relative entropy 
     $KL(\hat{\rho}_{n,n-1}||\rho_{n-1})$, for $n=1$ (a, c) and $n=2$ (b, d) 
     as a function of $1/\beta$. Purple curves refer to the unweighted collaboration network; yellow curves refer to the weighted version.}
     \label{fig::rel_ent_collab_projection}
\end{figure*}
In this section we reveal the information content encoded in the higher-order spectrum of the simplicial collaboration network by studying the higher-order spectral entropy and the relative entropy. A numerical analysis was carried out on the higher-order spectrum of both the weighted and the unweighted version of the simplicial collaboration network. In the unweighted simplicial complex all the affinity weights are set to one, i.e. $\omega_{\alpha}=1$ for every simplex $\alpha$.

In Figure \ref{fig::collab:VNentr} the Von Neumann entropy and its derivative are plotted as functions of ${1}/{\beta}$ in logarithmic scale. The continuous line represents the entropy, while the dotted line indicates the specific heat. The first row shows the results obtained for the unweighted network, and the second row shows those obtained for the weighted network. Starting from the left, the entropy of each order up to the order of the simplicial complex ($2$ in this case) is plotted. 
From the picture, a multi--scale behaviour of the entropy emerges from the observation of the entropy of order $0$ and $1$. This is highlighted by the presence of local minima/maxima in the derivative. More specifically, while a plateaux with very shallow local minima and maxima can be observed for order $0$ and $1$ in the unweighted network,  a more clear separation of time scales determined by two more pronounced peaks is observed for the weighted case.  These different time scales are  related to the meso-scale community structure of the collaboration simplicial complexes and its large scale topology. We note, however  that this separation of time scale is not apparent from the analysis of the spectral entropy of order $2$, and this is explained by the fact that  different communities typically are formed by triangles connected to triangles only through nodes, which does not allow any diffusion from triangles to triangles through links. 

Figure  \ref{fig::rel_entWvsUW} displays the  relative entropy between the weighted and unweighted version of the collaboration complex at each order. It emerges that the weights are responsible for changing significantly the diffusion properties at large time scales, as revealed by the larger values of the relative entropy for small values of $1/\beta$. 

Finally, in Figure \ref{fig::rel_ent_collab_projection}, the relative entropy between the network density of order $1$ and $2$ and their projections onto the $0-$chains and $1-$chains, are shown. The analysis is performed both on the weighted and the unweighted version of the simplicial complex as well. This last analysis reveals a very rich  structure of local maxima and minima when the spectral properties  of order $0$ and $1$ are compared, thus demonstrating the non-trivial effect of studying the higher-order diffusion properties of the simplicial complex. Indeed, while $\rho_0$ is only dependent on $L_{[0]}$, which captures the diffusion from nodes to links, $\rho_{1,0}$ is dependent on both $L_{[1]}^{down}$ and $L_{[1]}^{up}$, which describe diffusion from links to links through nodes and through triangles, respectively. Therefore the relative entropies $KL(\hat{\rho}_{1,0}||\rho_0)$ and its weighted version $KL(\hat{\rho}_{1,0}^{W}||\rho_0^W)$ are able to capture the effect of coupling diffusion processes occurring on different dimensions.
Since the considered collaboration dataset is a simplicial complexes of dimension $d=2$, when $\rho_1$ and $\hat{\rho}_{2,1}$ are compared we only see a decreasing behavior of the relative entropy with increasing values of $1/\beta$.

\section{Application to the Network Geometry with Flavor}
In this section we study the higher-order spectrum of the weighted simplicial complex model called ``Network Geometry with Flavor"  \cite{bianconi2015complex,bianconi2016network,bianconi2017emergent}, which is a very interesting textbench to validate our proposed methodology. The analysis follows the same steps of that carried out on the collaboration complex described in the previous section.

\subsection{Network Geometry with Flavor as a model of weighted simplicial complex}
\begin{figure*}
    \centering
    \includegraphics[width=0.8
    \textwidth]{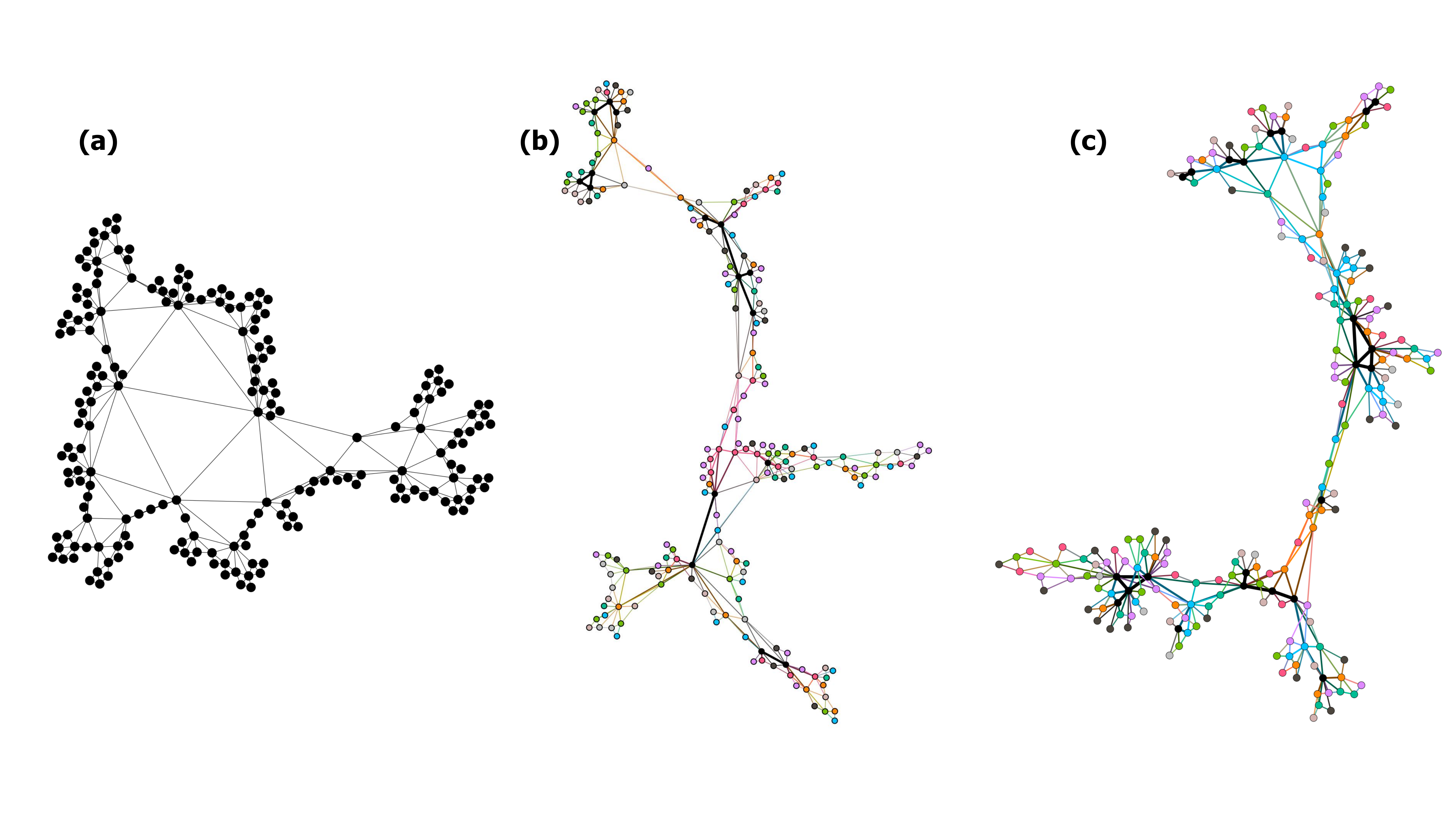}
\caption{Visualization of the network skeleton of the NGF with flavor $s=-1$, dimension $d=2$ and $\hat{\beta}=0$ (panel (a)), $\hat{\beta}=5$ (panel (b)), $\hat{\beta}=10$ (panel (c)). The color of the nodes and the links indicate their fitness value.}
\label{fig:ngf}
\end{figure*}

``Network Geometry with Flavor" (NGF) \cite{bianconi2015complex,bianconi2016network,bianconi2017emergent} is a model of weighted growing simplicial complexes which generates simplicial complexes whose network skeleton is small-world, modular, and hyperbolic.
The model is very comprehensive and general and admits several important extensions \cite{mulder2018network,courtney2017weighted,bianconi2021higher}. Here we focus on its original formulation for a $d$ dimensional simplicial complex in which each node $j$ is associated a feature called {\em energy} $\epsilon_j$ drawn from a distribution $g(\epsilon)$ that does not change in time.
The energy $\epsilon_{\alpha}$ associated to every simplex $\alpha$ of dimension $n>0$ is given by the sum of the energy of its nodes, i.e.
\bea
\epsilon_{\alpha}=\sum_{j\subset \alpha}\epsilon_j.
\eea
The {\em fitness} $\eta_{\alpha}$ of simplex $\alpha$  is given by 
\bea
\eta_{\alpha}=e^{-\hat{\beta} \epsilon_{\alpha}},
\eea
where $\hat{\beta}\geq 0$ is a tunable parameter of the model. (Note that for $\hat{\beta}=0$ all fitnesses are the same, i.e. $\eta_{\alpha}=1$ for any simplex $\alpha$).
Therefore for every simplex $\alpha$ of dimension $n>0$ they obey
\bea
\eta_{\alpha}=\prod_{j\subset \alpha}\eta_j,
\eea
where $\eta_j$ indicates the fitness associated to the generic node $j$ belonging to the simplex $\alpha$.
The NGF is a model for a growing $d$-dimensional simplicial complex  defined as follows. Starting at time $t=1$ from a single $d$-simplex, at each time a new $d$ dimensional simplex is added to the network. The new simplex has a single new node and it is attached to an existing $(d-1)$-dimensional face $\alpha$ chosen with probability 
\bea
\Pi_{\alpha}=\frac{\eta_{\alpha}(1-s+sk_{d,d-1}(\alpha))}{\hat{Z}},
\eea
where $s\in {-1,0,1}$ is a parameter called {\em flavor}, $k_{d,d-1}(\alpha)$ indicates how many simplices of dimension $d$ are incident already to the face $\alpha$, and  $\hat{Z}$ is the normalization constant. 
Here we focus on NGFs with flavor $s=-1$ and dimension $d=2$ with energy distribution $g(\epsilon)=1/10$ for $\epsilon\in \{0,1,2,3,4,5,6,7,8,9,10\}$. These are hyperbolic manifolds (random Farey graphs) that as a function of the parameter $\hat{\beta}$ undergo a topological transition from a small world network for low values of $\hat{\beta}$ to a finite dimensional network for large values of $\hat{\beta}$ \cite{bianconi2015complex}. In other words, the diameter of the network skeleton grows only logarithmically  with the network size for small values of $\hat{\beta}$, while it grows as a power of the network size for large values of $\hat{\beta}$ (see Figure $\ref{fig:ngf}$ for visualization of instance with $\hat{\beta}=0$ (small world topology) and with $\hat{\beta}=5,10$ (in the finite (Hausdorff) dimensional regime)).

\subsection{Higher-order entropy of Network Geometry with Flavor}
\label{subsec::VN_NGFexp}
 As the entropy can be interpreted as a quantity describing a mechanism of diffusion of information through the network, it is relevant to compare the behaviour of this quantity computed at different orders. 
We have considered  both an unweighted and a weighted version of the NGF model. By unweighted version of the NGF we indicate the case in which each simplex has bare affinity weights all equal to $1$, i.e. $\omega_{\alpha}=1$, for every simplex $\alpha$ of the simplicial complex. In other words, in the unweighted version of the NGFs the bare affinity weights are  independent from the fitness. On the contrary, the weighted version of the NGF indicates the case in which the bare affinity weights are given by the simplices fitness $\omega_{\alpha}=\eta_{\alpha}$.
In both the weighted and the unweighted version of NGFs the  topological weights are calculated according to Eq.(\ref{weigh:1}).

First, we calculated the higher-order spectral entropy and the higher-order specific heat of NGFs of flavor $s=-1$ and dimension $d=2$. This allows us to characterize the typical temporal scale of higher-order diffusion processes as a function of their order $n$.

\begin{figure*}[!htbp]
         \centering
        \includegraphics[width=\textwidth]{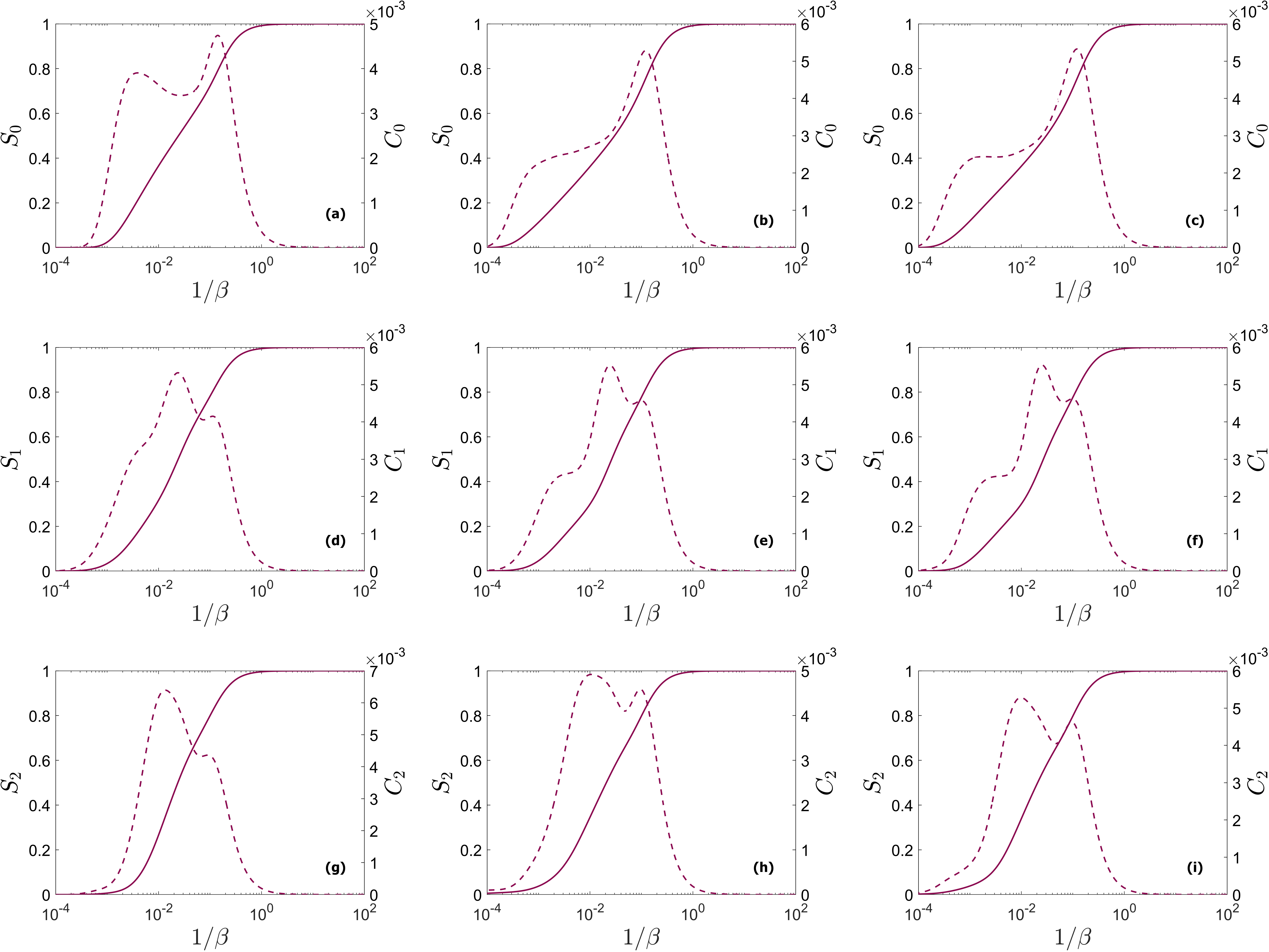}
        \caption{Entropy computed on the unweighted version of NGF with $200$ nodes and flavor $s=-1$ as a function of the reciprocal of the density parameter $1/\beta$ varying logarithmically in the interval $[10^{-4},10^{2}]$. From top to bottom, the Von Neumann entropy of order $0$ ((a), (b), (c)), $1$ ((d), (e), (f)) and $2$ ((g), (h), (i)) is represented with continuous lines. Dashed lines depict the derivative of the corresponding entropy with respect to the logarithm of the density parameter $\beta$. Each column corresponds to a different value of the model parameter $\beta$ (from left to right, $\beta=0,5,10$).}
         \label{fig:NGF_SC_unweighted}
\end{figure*}

\begin{figure*}[!htbp]
         \centering
         \includegraphics[width=0.8\textwidth]{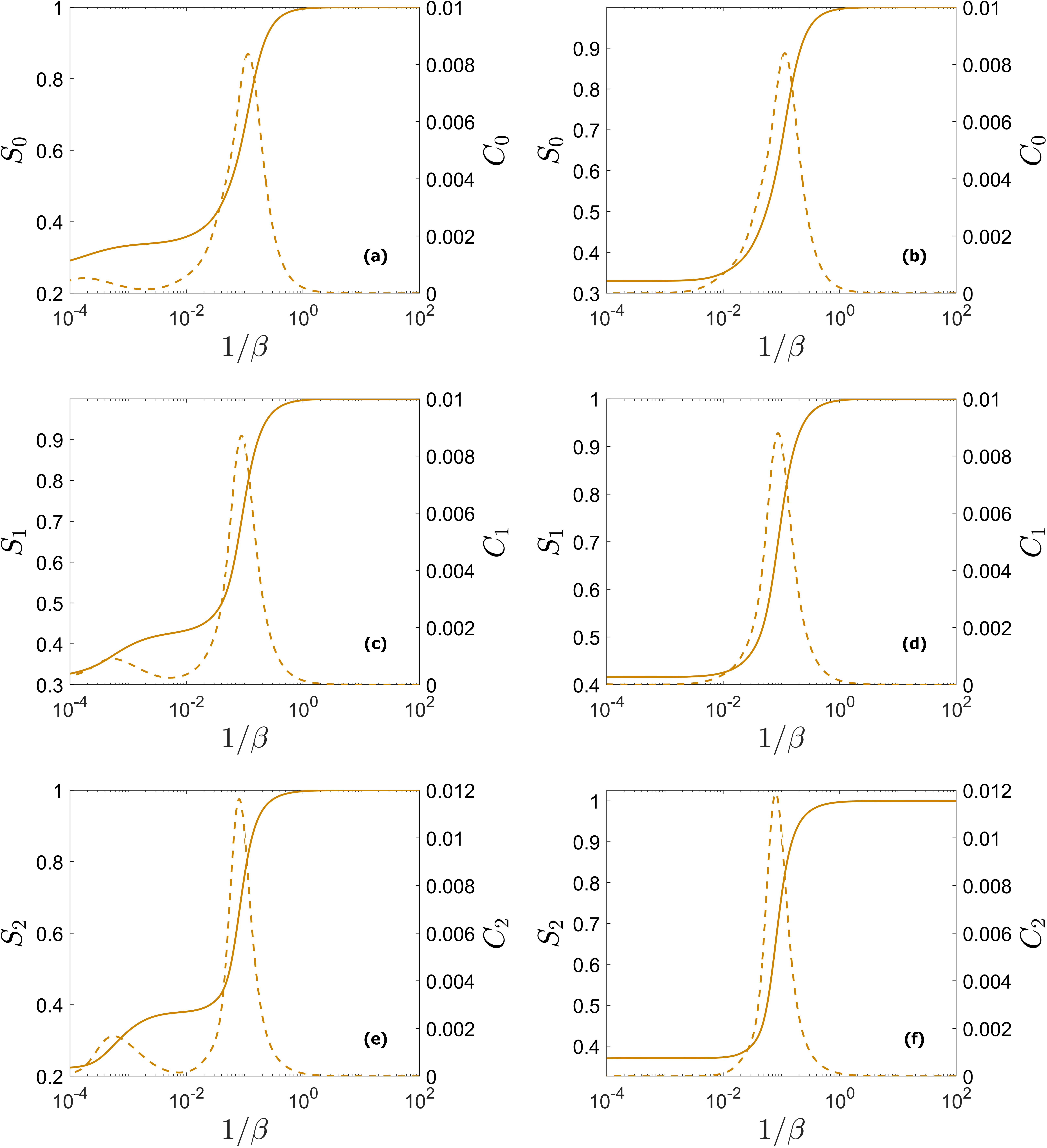}
        \caption{Entropy computed on the weighted version of NGF with $200$ nodes and flavor $s=-1$ as a function of the reciprocal of the density parameter $1/\beta$ varying logarithmically in the interval $[10^{-4},10^{2}]$. From top to bottom, the Von Neumann entropy of order $0$ ((a), (b)), $1$ ((c), (d)), and $2$ ((e),(f)) is represented with continuous lines. Dashed lines depict the derivative of the corresponding entropy with respect to the logarithm of the density parameter $\beta$. Each column corresponds to a different value of the model parameter $\beta$ (from left to right, $\beta=5,10$).}
         \label{fig:NGF_SC_weighted}
\end{figure*}

 Figure \ref{fig:NGF_SC_unweighted}  shows the higher-order spectral  entropy of order $0, 1$ and $2$ and the corresponding higher-order specific heat  as functions of $1/\beta$ for the unweighted NGF with $s=-1,d=2$ and  $\hat{\beta}\in \{0,5,10\}$; Figure \ref{fig:NGF_SC_weighted} refers to the weighted NGF plotted here only for the relevance cases $\hat{\beta}\in \{5,10\}$. 
 All the simulations are performed on simplicial complexes of $N=200$ nodes. Moreover, results from Fig.\ref{fig:NGF_SC_unweighted} and \ref{fig:NGF_SC_weighted} were obtained by averaging the results obtained over 100 realizations of the NGFs.
 From Figures  \ref{fig:NGF_SC_unweighted} and  \ref{fig:NGF_SC_weighted} we clearly notice that diffusion at different order can have different temporal scales, indicated by the local maxima and minima of the higher-order specific heat. Moreover, it emerges that taking into account the fitness of the simplices as the bare affinity weights of the diffusion dynamics, clearly allows to establish characteristic scales of the higher-order diffusion dynamics that are more clearly distinguished and well defined, since the peaks of the specific heat are more narrow in the weighted case.

\begin{figure*}[!htbp]
         \centering
        \includegraphics[width=0.95\textwidth]{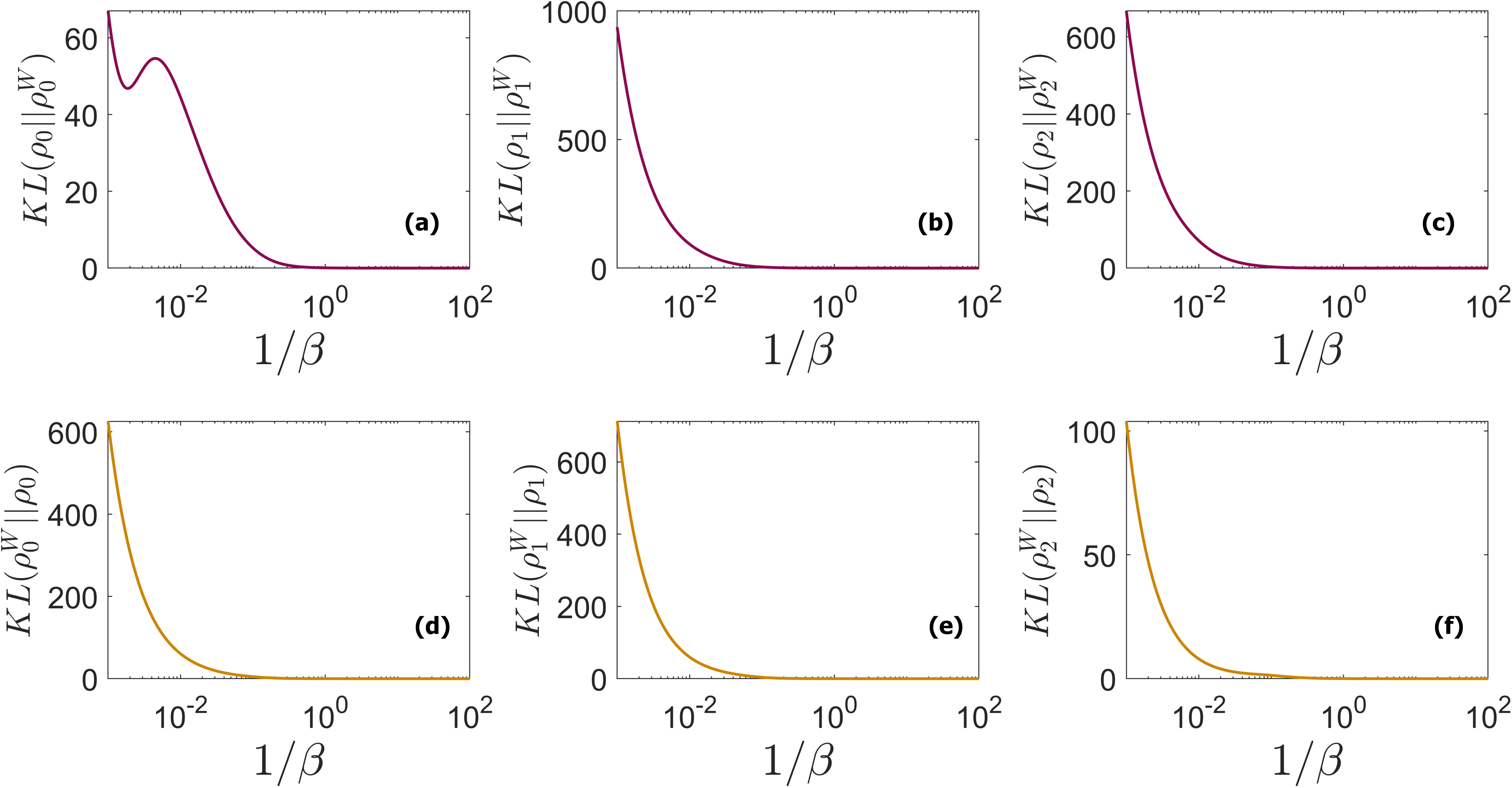}
        \caption{Quantum relative entropy between weighted and unweighted NGF complex having $200$ nodes, flavor $s=-1$ and $\hat{\beta}=5$ as a function of the reciprocal of the density parameter $1/\beta$ varying logarithmically in the interval $[10^{-3},10^{2}]$.}
         \label{fig::rel_WvsUW_NGF}
\end{figure*}

{Fig. \ref{fig::rel_WvsUW_NGF} depicts the quantum relative entropy between the weighted and the unweighted version of the NGF simplicial complex for $\hat{\beta}=5$. This quantity generally follows a decreasing behaviour, reflecting the fact that the weights impact significantly on the diffusion properties of the complex at large time scales. Interestingly, at order $0$, Fig. \ref{fig::rel_WvsUW_NGF} (a) highlights that the quantum relative entropy is not monotonic, but admits a local minimum and a local maximum.}

\begin{figure*}[!htbp]
         \centering
         \includegraphics[width=\textwidth]{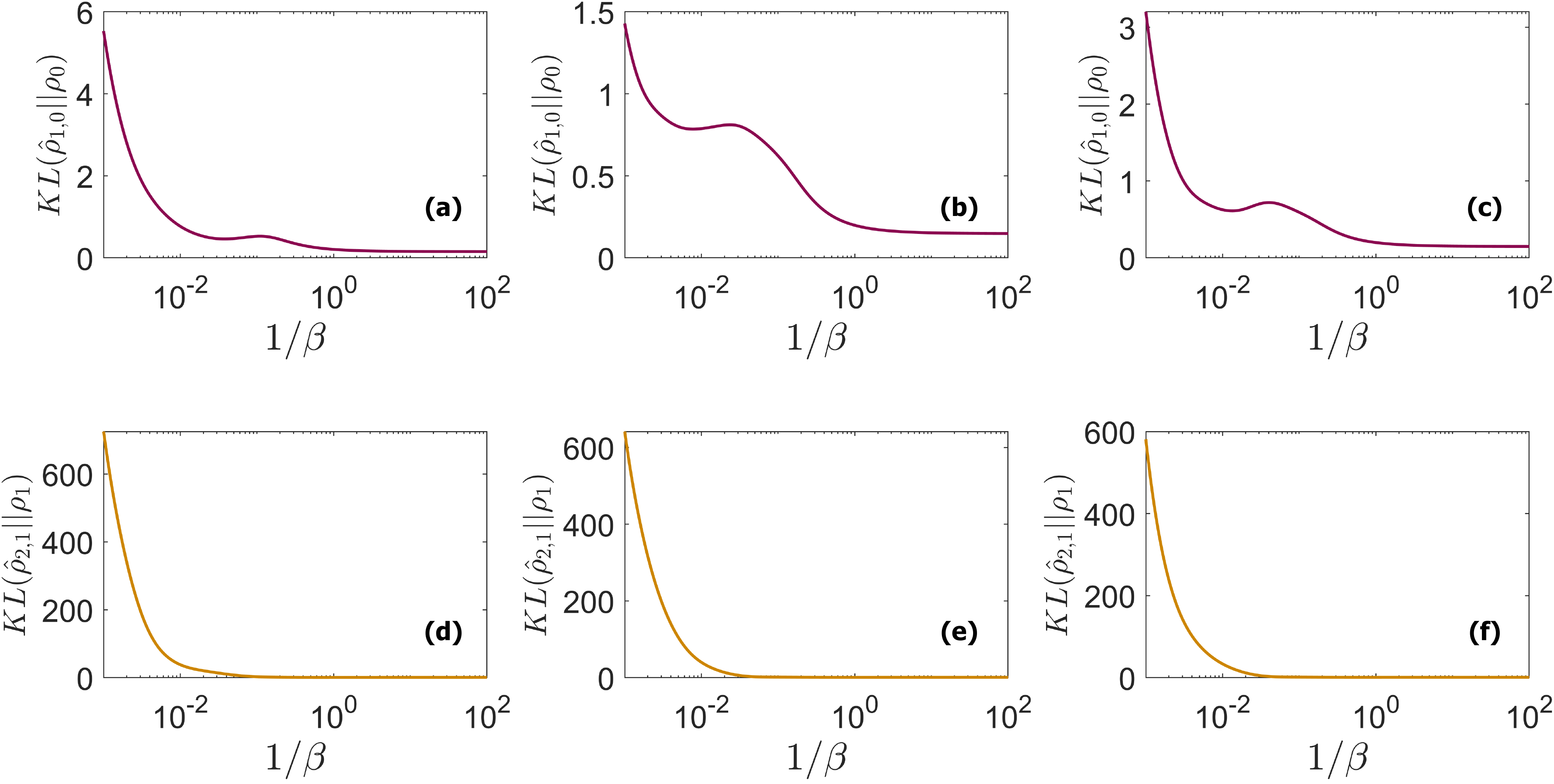}
        \caption{Relative entropy $KL(\hat{\rho}_n||\rho_{n-1})$  computed on the unweighted version of NGF with 200 nodes, flavor $=-1$, as a function of $1/\beta$, where $\beta$ is the density parameter. Starting from the left, (a) and (d) correspond to the model parameter $\beta=0$, (b) and (e) to $\beta=5$, (c) and (f) to $\beta=10$.}
         \label{fig:re_NGF_unweighted}
\end{figure*}

\begin{figure*}[!ht]
         \centering
         \includegraphics[width=0.7\textwidth]{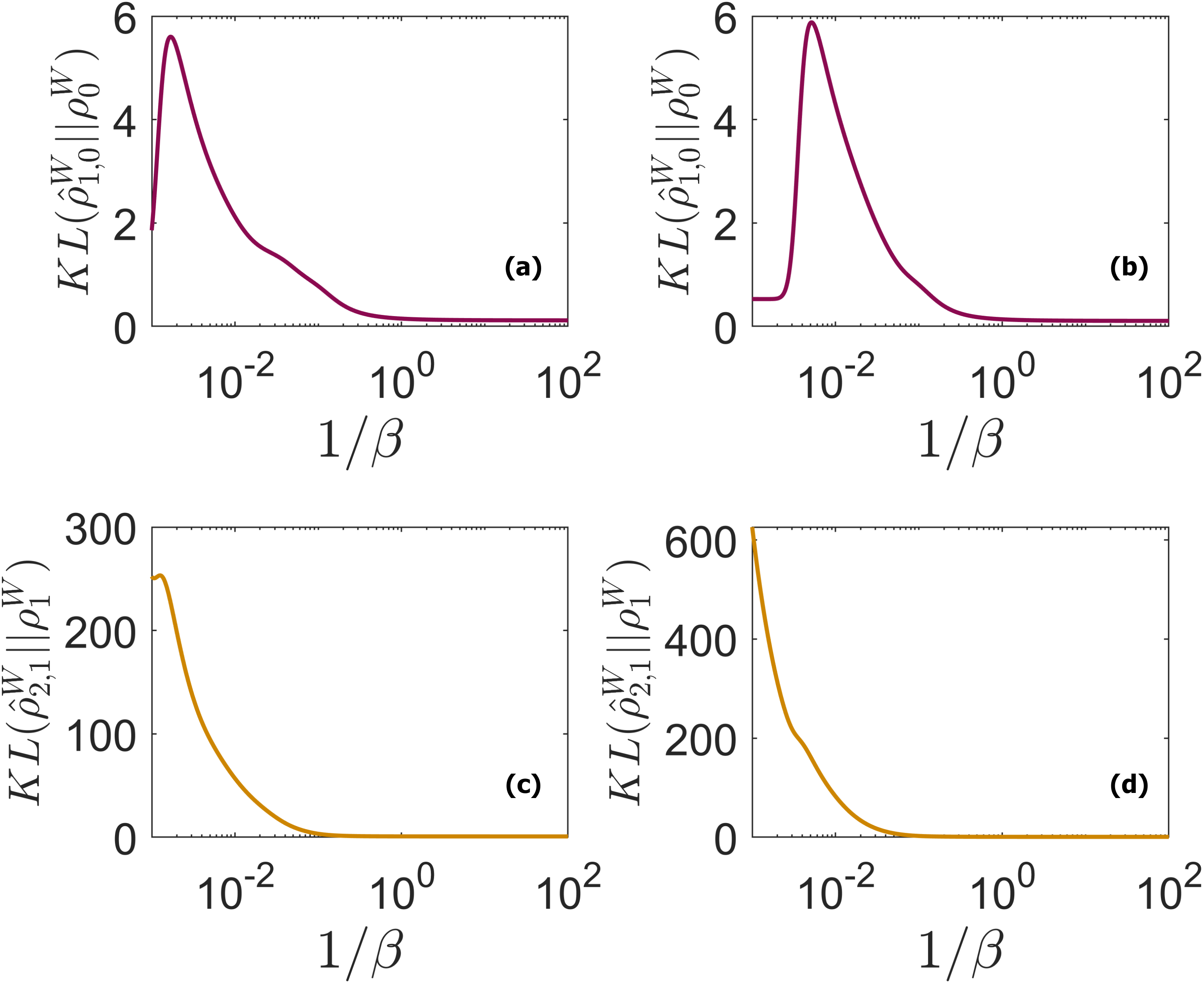}
        \caption{Relative entropy $KL(\hat{\rho}_n||\rho_{n-1})$  computed on the weighted version of NGF with 200 nodes, flavor $=-1$, as a function of $1/\beta$, where $\beta$ is the density parameter.(a) and (c) correspond to the model parameter $\beta=5$, (b) and (d) to $\beta=10$.}
         \label{fig:re_NGF_weighted}
\end{figure*}

In Figures $\ref{fig:re_NGF_unweighted}$ and  $\ref{fig:re_NGF_weighted}$ we plot the quantum relative entropy calculated respectively on the unweighted and weighted version of the NGF with flavor $s=-1$, dimension $d=2$ and $\hat{\beta}\in \{0,5,10\}$ (for the unweighted case) and $\hat{\beta}\in \{5,10\}$ (for the weighted case).
Interestingly the effect of changing $\hat{\beta}$ is noticeable from the study of the quantum relative entropy. Indeed  the quantum relative entropy $KL(\hat{\rho}_{1,0}||\rho_0)$ develops a non monotonicity for $\hat{\beta}=5,10$ in the unweighted case, and a very well defined peak for the weighted case.

\section{Conclusions}
Here we propose to study weighted simplicial complexes with a precise convention for the weights of the simplices to overcome the limitation of simplicial complexes to capture arbitrary higher-order network data.
The advantage of the proposed mathematical framework is that the weighted simplicial complexes have a rich higher-order structure that can be probed with higher-order weighted and normalized Hodge Laplacians.
The spectrum of the higher-order Hodge Laplacians allows to use tools of information theory to quantify the information content included in the higher-order spectrum of the simplicial complex and the properties of higher-order diffusion processes. Indeed, we propose the notion of higher-order spectral entropy and we show that this quantity can be used to  characterize the typical temporal scales of higher-order diffusion. Moreover, the higher-order relative spectral entropy  allows us to compare the information content encoded in the spectrum of Hodge Laplacians of different dimensions.

The proposed approach is here tested on a real higher-order collaboration dataset that is extracted from bibliometric data by adopting a procedure to weight the higher-order collaborations that is based on an extension of a widely used convention adopted for simple networks.

Finally the approach is also applied to the weighted version of the simplicial complex model ``Network Geometry with Flavor". The analysis reveals a the dependence of the higher-order diffusion properties on the simplicial complex as a function of a control parameter. This give an insight on how the higher-order spectral properties of the simplicial complex depend on its underlying topology.  

We believe that the proposed choice of weights for simplicial complexes, and the associated normalized Hodge Laplacian, will constitute a very useful tool to capture the structure of higher-order network data. In addition, since Hodge Laplacians are increasingly used to capture the dynamics of topological signals on simplicial complexes, we believe that the proposed normalized and weighted Hodge Laplacians would be a very useful tool to describe the dynamics of topological signals on weighted simplicial complexes.

\bibliographystyle{apsrev4-1}
\bibliography{entropy_references}
\appendix
\section{Normalized higher-order Laplacians}
\label{AppendixA}
\label{Ap:normalization}
Thanks to the Hodge decomposition of $L_{[n]}$ the non-zero eigenvalues of $L_{[n]}$ are either non-zero eigenvalues of $L_{[n]}^{down}$ or non-zero eigenvalues of $L_{[n]}^{up}$.
Moreover, we note that $L_{[n]}^{up}$ and $L_{[n+1]}^{down}$ are isospectral, i.e. they have the same non-zero eigenvalues.
It follows that to demonstrate that for a given choice of $a_{n}$ and the metric matrices $G_{[n]}$ all the higher-order Laplacians $L_{[n]}$ are normalized, it can be equivalently demonstrated that under these conditions every higher-order up Laplacian $L_{[n]}^{up}$ is normalized, independently on the value of $n$ with $0\leq n<d$.
In order to do that we first notice that $L_{[n]}^{up}$ given by 
\bea
L_{[n]}^{up}=G_{[n]}\bar{B}^*_{[n+1]}G_{[n+1]}^{-1}
\bar{B}_{[n+1]}
\eea is isospectral to the symmetrically normalized higher order Laplacian $L_{[n]}^{up,symm}$ given by 
\bea
L_{[n]}^{up,symm}=G_{[n]}^{1/2}\bar{B}^*_{[n+1]}G_{[n+1]}^{-1}\bar{B}_{[n+1]}G_{[n]}^{1/2}.
\eea
The spectrum of $L_{[n]}^{up,symm}$ can be studied by considering  the Rayleigh quotient \cite{chung1997spectral}. Given a generic  column vector $\bm{X}$ defined on the $n$-simplices of the considered simplicial complex, then the Rayleigh quotient of the symmetrically normalised higher order Laplacian is given by 
 \bea\label{raylnorm}
      RQ({L}_{[n]}^{up,sym},\bm X)=\frac{\bm{X}^T{L}_{[n]}^{up,symm}\bm{X}}{\bm{X}^T\bm{X}},
      \eea
      and we have for every eigenvalues $\lambda$ of $L_{[n]}^{up,symm}$
      \bea
      0=\min_{\bm X} RQ({L}_{[n]}^{up,sym},\bm X)\geq\lambda\leq\max_{\bm X} RQ({L}_{[n]}^{up,sym},\bm X).
      \eea
      Let us use the following notation.
     With  $w_{\tau(r,s)}$ is the topological weight of the $n+1$ simplex $\tau(r,s)$ incident to a chosen pair of $n$-simplices $r$ and $s$ that are up adjacent. With $s_r$ we indicate the sum of the weights of the $n+1$ simplices incident to the $n$-simplex $r$, i.e.
     \bea
     s_r=\sum_{\tau\supset r}w_{\tau}.
     \eea
     We the introduce the $N_{[n]}\times N_{[n]}$ matrices $A_{\updownarrows}^{[n]}$ and $A_{\upuparrows}^{[n]}$.
     The  matrix $A_{\updownarrows}^{[n]}$ has  elements $A_{\updownarrows}^{[n]}(r,s)=1$ if the $n$-simplex $r$ is up adjacent to the $n$-simplex $s$ and $r$ and $s$ have the opposite orientation with respect to the common ($n+1$) simplex $\tau(r,s)$; otherwise  $a_{\updownarrows}^{[n]}(r,s)=0$.
     Similarly, $A_{\upuparrows}^{[n]}$ has elements $A_{\upuparrows}^{[n]}(r,s)=1$ if an only if $r$ and $s$ are up adjacent with the same orientation with respect to $\tau(r,s)$ and otherwise has elements $a_{\upuparrows}^{[n]}(r,s)=0$.
 Note that given the definition of the matrices $A_{\updownarrows}$ and $A_{\upuparrows}$, for any choice of a pair of simplices $r$ and $s$ we cannot have simultaneously $A_{\updownarrows}^{[n]}(r,s)=1$ and $A_{\upuparrows}^{[n]}(r,s)=1$, therefore we have
\bea 
|A_{\updownarrows}^{[n]}(r,s)-A_{\upuparrows}^{[n]}(r,s)|\leq 1.
\label{dis1}
\eea
 
      By setting $\bm{Y}=G_{[n]}^{1/2}\bm{X}$ the Rayleigh quotient can be expressed as 
      \bea
      RQ({L}_{[n]}^{up,sym},\bm X)=\frac{\bm{Y}^T \hat{L}_{[n]}^{up} \bm{Y}}{\bm{Y}^T G_{[n]}^{-1} \bm{Y}},
      \eea
 where $\hat{L}_{[n]}^{up}$ is the $N_{[n]}\times N_{[n]}$ matrix of elements
   \bea
 \hspace*{-6mm}\hat{L}_{[n]}^{up}(r,s)=\frac{1}{n+2}\left[s_r^{[n]}\tilde\delta_{r,s}-w_{\tau(r,s)}\left(A_{\updownarrows}^{[n]}(r,s)-A_{\upuparrows}^{[n]}(r,s)\right)\right].
 \eea
We  can easily show that 
 \bea \label{dis}
 \bm{Y}^{\top}\hat{L}_{[n]}\bm{Y}\leq \sum_{r}s_rY_r^2.
 \eea
 Indeed, by using Eq. (\ref{dis1}) and the inequality $2|xy|\leq x^2+y^2$ we obtain
 \bea
 \left|\left(A_{\updownarrows}^{[n]}(r,s)-A_{\upuparrows}^{[n]}(r,s)\right)Y_rY_s\right|\leq Y_rY_s\leq \frac{1}{2}[Y_r^2+Y_s^2].
 \eea
 Therefore  we obtain 
 \bea \label{Yr}
  \bm{Y}^{\top}\hat{L}_{[n]}\bm{Y}&=&\sum_{r,s}Y_r\hat{L}^{up}_{[n]}Y_s\nonumber \\
  &\leq& \frac{1}{n+2}\left(\sum_{r}s_rY_r^2+\sum_{r,s}w_{\tau(r,s)}Y_r^2\right).
 \eea
 Now we note that since every $n+1$-simplex $\tau$ incident to $r$ is also incident to other $n+1$ $n$-simplices,  we obtain
 \bea
 \sum_{s}w_{\tau(r,s)}=(n+1)s_r.
 \eea
 Using this relation in Eq. (\ref{Yr}) Eq. (\ref{dis}) follows directly.
 
 Finally, using the definition of the matrix $G_{[n]}^{-1}$ given by Eq.(\ref{Gm0}) with topological weights given by Eq.(\ref{weigh:1}), since  the bare affinity weight satisfy  $\omega_{\alpha}\geq 0$ for every simplex $\alpha$, it follows immediately that
 \bea \label{disG}
 \bm Y^{\top}G_{[n]}^{-1}\bm Y\leq \sum_r s_rY_r^2.
 \eea
 Therefore, using Eq.(\ref{dis}) and Eq.(\ref{disG}) we obtain  that the Rayleigh quotient, and hence the spectrum of $L_{[n]}^{up,symm}$, is bounded by one.
 As a consequence of this, the eigenvalues of $\lambda$ of $L_{[n]}^{up}$ satisfy 
 \bea
 0\leq \lambda\leq 1.
 \eea
 This concludes the proof that the spectrum of the higher-order Laplacian $L_{[n]}$ is normalized and has an upper bound one, for any order $n$, provided that the topological weights are chosen according to Eq.(\ref{weigh:1}).

\end{document}